# Machine-Type Communication Waveforms: An Exploration of New Dimensions

Michael Wang[1] and L. Wang[2], and X. You[1], *Fellow, IEEE*

*Abstract*—This paper derives a generalized class of waveforms with an application to machine-type communication (MTC) while studying its underlying structural characteristics in relation to conventional modulation waveforms. First, a canonical waveform of frequency-error tolerance is identified for a unified preamble and traffic signal design, ideal for MTC use as a composite waveform, commonly known as a transmission burst. It is shown that the most widely used modulation schemes for traffic signals, e.g., FSK and LoRa modulation, are simply *subsets* of the canonical waveform. The intrinsic characteristics and degrees of freedom the waveform offers are then explored. Most significantly, a new waveform dimension is uncovered and exploited as additional degrees of freedom for satisfying the MTC requirements, i.e., energy and resource efficiency and robustness. The corresponding benefits are evaluated analytically and numerically in AWGN, frequency-flat, and selective channels. We demonstrate that neither FSK nor LoRa can fully address the mIoT requirements since neither fully exploits the degrees of freedom from the perspective of the generalized waveform class. Finally, a solution is devised to optimize energy and resource efficiency under various deployment environments and practical constraints while maintaining the low-complexity property.

## I. INTRODUCTION

The Internet of Things (IoT) is a worldwide infrastructure that facilitates advanced services by seamlessly integrating "things" into the information network. Massively distributed sensing creates interactions among the environment, humans, and real and virtual objects. This poses unprecedented challenges to IoT systems in terms of massive machine communication, commonly known as *machine-type communication* (MTC), which serves as the backbone of IoT, providing connectivity to a network of IoT end devices.

There are myriad types of IoT networks, among which is the most popular *sensor network*, typically consisting of a massive number of sensing devices [1], commonly referred to as the massive IoT (mIoT) network [2]-[4]. MTC for mIoT is subsequently referred to as mMTC. Nevertheless, the attributive "massive" is more *symbolic* than literal and is commonly used in the jargon of the IoT community to represent a specific type of IoT with the following characteristics.

Indeed, mMTC differs starkly from human-type communication (HTC). HTC (e.g., the 4G-LTE) is characterized by high data rate and low latency, empowered by wide bandwidth, high-cost-and-performance hardware, and high-capacity rechargeable batteries, and fueled by a large market and ever-increasing bandwidth-hungry multimedia applications. On the contrary, mIoT consists of resource-constrained, small-form-factored, mass-producible, and disposable devices (typically, sensors) and is dominated by *short-burst data services* on the *uplink* (i.e., from the device to the sink or server, where the uploaded data are utilized by the server application). For instance, MTC allows the development of mIoT solutions that use distributed sensors for large-area coverage. Through dispersed sensors, mIoT in forestry and environmental science enables real-time monitoring of the environment.

Just like everything else in the universe, communication requires energy. Connecting IoT end devices to a mains supply can be challenging, especially for low-cost mIoT deployments such as sensor networks. These deployments often involve devices with limited resources and small form factors and are mass-producible and disposable. This implies three things: 1) the device must be battery-powered, 2) the radio-frequency (RF) frontend is of low quality, which results in low transmit power, high error vector magnitude (EVM), and no GPS or similar synchronization sources, and 3) low processing power. Both 1) and 3) speak of a constraint on device complexity. Since the battery's energy must be used at its optimum to ensure longevity (e.g., years), the low-energy mMTC technology becomes a key factor of practical mIoT. Meanwhile, as the name implies, massive IoT involves supporting potentially a large number of IoT devices, which puts a high demand on the transceiver's resource efficiency. The volume of information that can be transferred and retrieved per unit of consumed energy and communication resources (time and frequency in degrees of freedom or DoF), i.e., *energy and resource efficiency*, should be maintained as high as possible in the end devices, in other words, the amount of energy and resource consumed per information bit should be as low as possible.

Apparently, connecting mIoT devices by running wires to the service network can often be impractical—just as powering the devices via cabling is impractical. This means that connectivity (at least the "last mile") must be provided wirelessly, i.e., between IoT devices and the control station in a cellular infrastructure-based wide-area MTC network [5]. Nonetheless, a wireless channel is known for its unreliability, mainly caused by fading due to multipath that degrades communication channel conditions and reliability [6].

Although the specific requirements of the mIoT device depend on the technology and vary with the application and deployment environment, they can be generally categorized in four *aspects*: 1) high energy efficiency, which affects the network's lifespan; 2) high resource efficiency, which determines the number of devices the network can support with given system communication resources; 3) robustness to channel impairments since mIoT networks are deployed in wireless environments; 4) low complexity and manufacturing cost.

For example, the device battery life in a typical forest monitoring application requires a minimum battery life of ten years under normal operating conditions (e.g., generating four reports daily). Given the reporting message size, deployment channel condition, and battery capacity (constrained by the cost and form factor), it imposes a corresponding requirement on the device's energy efficiency. Similarly, resource efficiency affects the device density the network can support for a given bandwidth. Therefore, the *degree of the fulfillment* of these four characteristics or requirements of the mIoT devices directly affects the competitiveness of the technology in the IoT market.

Compared with human-centric communications, the mMTC *traffic pattern* shifts from large to small packets, high to low data rates, and downlink to uplink. Nevertheless, despite the low data rates, the network can be deeply strained by potentially large volumes of concurrent accessing devices under a limited frequency bandwidth. Therefore, mMTC represents a new wireless communication paradigm, requiring a different set of technologies to those designed to accommodate human-centric communication traffic.

In this paper, by mIoT or mMTC, we mean the system or technology with these device characteristics and traffic patterns.

All these characteristics or requirements constrain the practical design of mMTC transceivers. One major challenge is the mMTC modulation waveform design—a non-trivial task since resource efficiency, energy efficiency, communication reliability, and device complexity go against each other: higher resource efficiency and higher reliability typically incur higher complexity and lower energy efficiency.

The state-of-the-art Narrow-Band IoT (NB-IoT) belongs to the *high-end*, low-power, and wide-coverage MTC technology [7]-[10]. Since it is built on top of the 4G-LTE network originally developed for HTC, it essentially inherits the LTE framework, including network configuration, layer protocols (e.g., the network and medium access protocols), and sophisticated physical waveforms: single carrier frequency division multiplexing (SC-FDM) with QPSK modulation combined with powerful turbo coding and rate adaptation. Furthermore, repetition coding is employed to boost the processing gain under a narrowband (i.e., 3.75 kHz) configuration to extend the transmission range, enabling low-power transmission with high resource efficiency. While turbo coding can provide significant coding gains that benefit resource and energy efficiency, it comes with increased complexity, manufacturing costs, and operating energy. As a result, the energy benefit gained from the coding gain may be offset. Moreover, QPSK is not particularly suited for mMTC due to phase discontinuity (more details later). In that sense, NB-IoT is a low-power and high-resource-efficiency MTC technology, not particularly low-energy or low-cost.

On the other end (the *lower end*), LoRa modulation is a different kind of mMTC technology [11][12]. Due to its uncontested low cost and low energy characteristics, it has become a mainstream mMTC technology. As a proprietary "black technology," limited information is available to the public, although some excellent studies and reports on LoRa modulation have recently appeared in the literature. In [13] and [14], the LoRa modulation waveform is mathematically shown to be a memoryless, continuous phase modulation and can be approximated as an orthogonal modulation for large modulation orders. [15] provides a theoretical derivation of an optimal receiver entailing a low-complexity demodulation process by the Fast Fourier Transform. Closed-form approximations of bit error rate performance are derived in [16] and [17] for both additive white Gaussian noise (AWGN) and Rayleigh fading channels. Performance under other channel conditions can be found in [18]-[20], indicating that LoRa modulation only minimizes the power amplifier efficiency, just like most legacy frequency modulations do. The ultimate required energy of the LoRa modulation in terms of the required received energy per information bit (i.e., the receive $E_b/N_0$ ) is higher than the legacy frequency-shift keying (FSK) under conditions like line-of-sight and frequency-flat channels—typical for narrowband MTC [21]. Moreover, the resource efficiency is inherently limited by its modulation order and reaches its maximum at the minimum modulation order, whereas other linear modulation techniques like quadrature amplitude modulation (QAM) are theoretically unlimited.

Inspired by the energy efficiency of orthogonal modulation (e.g., FSK) and the resource efficiency of linear modulation (QAM), [22] combines FSK with QAM, termed frequency and quadrature-amplitude modulation (FQAM). Different from pure FSK, the active frequency of an FSK symbol is further modulated by QAM, increasing the spectral efficiency compared to pure FSK. A similar scheme can be found in [23], in which the author combines the LoRa modulation with PSK (PSK-LoRa) to overcome the resource inefficiency of LoRa modulation. Nevertheless, both FQAM and PSK-LoRa suffer from the phase discontinuity introduced by QAM or PSK, which causes spectrum regrowth and decreases the transmit symbol rate compared with the original FSK and LoRa modulation. It hikes up the peak-to-average power ratio (PAPR) and degrades energy efficiency if spectrum shaping is applied to improve the frequency localization or spectral confinement.

The main contribution of this paper is the derivation of a generalized class of modulation waveforms for mMTC with the four characteristics mentioned, which includes a new dimension that has not been explored as additional degrees of freedom for addressing the mMTC challenges. Specifically, we show that the modulation schemes mentioned above, i.e., QAM, FSK, LoRa, and their variants (FQAM and PSK-LoRa), are all special cases of this derived class of mMTC waveforms. We first analytically examine this class of waveforms' intrinsic structural characteristics and explore the potential degrees of freedom for mMTC. We demonstrate that not all degrees of freedom are particularly suitable for mMTC, which excludes some of these modulation techniques from

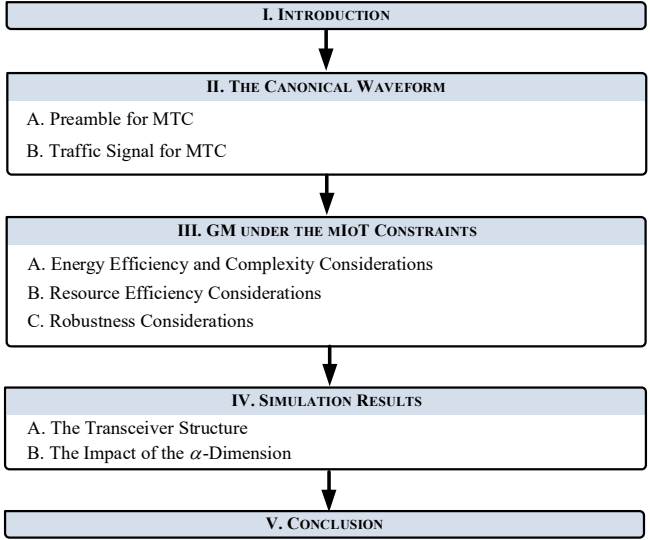

```
I. INTRODUCTION
        │
        ▼
II. THE CANONICAL WAVEFORM
    A. Preamble for MTC
    B. Traffic Signal for MTC
        │
        ▼
III. GM UNDER THE mIoT CONSTRAINTS
    A. Energy and Complexity Considerations
    B. Resource Efficiency Considerations
    C. Robustness Considerations
        │
        ▼
IV. SIMULATION RESULTS
    A. The Transceiver Structure
    B. The Impact of the α-Dimension
        │
        ▼
V. CONCLUSION
```

Figure 1 Organization diagram to show the structure of this paper.

TABLE I
LIST OF MAIN SYMBOLS DEFINED IN THIS PAPER.

| | |
|---|---|
| GM | Generalized modulation. |
| mGM | Generalized modulation for mMTC. |
| DoF | Degree(s) of freedom. |
| PAPR | Peak-to-average power ratio |
| $\eta$ | Resource consumption per information bit (DoF/bit). |
| $\varepsilon$ | (Normalized) energy consumption per information bit (dB/bit). |
| $\alpha, \beta, \rho$ | GM parameters defined in (16). |
| $\mathcal{S}^{\mathrm{GM}}$ | GM symbol set defined in (18). |
| $\mathcal{A}^{\mathrm{GM}}$ | GM $\alpha$-domain defined in (30). |
| $\mathcal{B}^{\mathrm{GM}}$ | GM $\beta$-domain defined in (26). |
| $\mathcal{C}^{\mathrm{GM}}$ | GM $\rho$-domain defined in (23). |
| $\mathcal{S}^{\mathrm{mGM}}$ | mGM symbol set defined in (57). |
| $\mathcal{A}^{\mathrm{mGM}}$ | mGM $\alpha$-domain defined in (56). |
| $\mathcal{C}^{\mathrm{mGM}}$ | mGM $\rho$-domain defined in (52). |

mMTC. Nevertheless, a new waveform dimension is identified and exploited as additional degrees of freedom for resource, energy efficiency, and robustness while maintaining low complexity, similar to LoRa modulation.

As diagrammed in Figure 1, the remainder of the paper is organized as follows: Section II first introduces a canonical waveform suitable for the preamble signal to an mMTC transmission burst and then sets out the main theme for this paper, i.e., a generalized class of modulation waveforms derived from the canonical waveform for the traffic bearer of the burst. As the main focus of this paper, Section III discusses the *generalized modulation* (GM) waveform design under the mMTC constraints, studies its structural properties, explores the embedded degrees of freedom, and evaluates their potential benefits and associated costs concerning energy, implementation complexity, resource efficiency, and robustness against wireless channels. Section IV applies GM for mMTC and numerically evaluates the performance under practical wireless channel conditions. Section V concludes the paper.

In this paper, we represent the amount of consumed energy per successfully received information bit as $\varepsilon$ (dB/bit), where the energy is normalized to the noise power spectral density at the receiver,

$$\varepsilon = E_{\mathrm{b}}/N_0, \tag{1}$$

where $E_{\mathrm{b}}$ is the bit energy, and $N_0$ is the noise power spectral density. Under this well-known convention (i.e., bit SNR), the energy is represented in dB (unit*less*) rather than Joules. Apparently, (1) is the inverse of energy efficiency, i.e., *energy inefficiency*, to be exact. Hence, a smaller $\varepsilon$ means greater energy efficiency.

Similarly, we denote the consumed communication resources (DoF) per successfully received information bit as $\eta$ (DoF/bit), i.e., the inverse of *resource efficiency*,

$$u \triangleq \eta^{-1} \text{(bits/DoF)}, \tag{2}$$

which is essentially *resource inefficiency* to align with the definition of $\varepsilon$ (energy inefficiency). Obviously, the less the value of $\eta$ is, the greater the resource efficiency.

The actual energy and resource efficiency characteristics depend on the design of an mMTC transceiver but are ultimately constrained by the Shannon-Hartley formula [24]

$$\varepsilon = \eta \cdot (2^{1/\eta} - 1). \tag{3}$$

The symbol definitions are tabulated in TABLE I for easy reference.

## II. THE CANONICAL WAVEFORM

In wireless communication, a transmitter and a receiver must rendezvous at the same time, frequency, and space before the communication can start. A typical mMTC physical *transmission burst* for carrying upper layer traffic (e.g., application data) from the sender to the receiver thus consists of a *preamble signal* followed by a *traffic signal*, as illustrated in Figure 2. The preamble is used for a receiver to detect the incoming burst and then time-and-frequency synchronize to the burst, and the traffic signal carries the payload of the burst—upper layer traffic.

This section develops a unified or "canonical" waveform suitable for the preamble and traffic signals of an mMTC transmission burst. This unified waveform design favors a low-complexity transceiver, which is crucial for mMTC. We start with analyzing how the transceiver synchronization error affects the burst detection performance and derive a canonical waveform intrinsically resilient to such errors. We then exploit this waveform for bearing traffic to obtain a generalized modulation or GM waveform set.

### A. Preamble for MTC

Although it is quite common in the literature to assume perfect synchronization between transceivers, in reality, it is a non-trivial task to detect a transmission burst and establish synchronization between communicating peers, especially for complexity- and energy-constrained mMTC transceivers with low-stability synchronization sources. The intrinsic frequency error of a low-cost and low-energy IoT device could be well above $\pm 5$ ppm (e.g., $\pm 20$ ppm) [25] [26], especially under a large ambient temperature variation due to 1) the lack of temperature-controlled crystal oscillator (costly and energy-

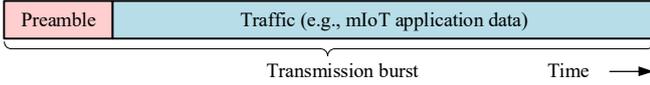

| Preamble | Traffic (e.g., mIoT application data) |

Transmission burst ————————————— Time →

Figure 2 Massive MTC transmission burst structure—a composite waveform consisting of a preamble and a traffic bearer.

consuming)—the luxury an mMTC transceiver cannot afford and 2) after a long deep sleep—typical for short-burst-data mIoT services.

Let us briefly review how the frequency error affects the detection performance. Assuming two communicating peers communicating via the RF signal,

$$\tilde{x}(t) = x(t) \cdot e^{j2\pi f_c t}, \quad (4)$$

where $x(t)$ is the baseband signal, $t$ (sec) denotes time, and $f_c$ (Hz) represents the carrier frequency (in the RF band). The carrier frequency offset between the transmitter and the receiver is $\Delta f$ (Hz) due to, e.g., a frequency error of the local oscillator of the transmitter, the received RF signal can be expressed as

$$x(t) \cdot e^{j2\pi(f_c + \Delta f)t} = \left( x(t) \cdot e^{j2\pi\Delta f t} \right) \cdot e^{j2\pi f_c t}. \quad (5)$$

The corresponding baseband signal after down-conversion is then

$$x(t) \cdot e^{j2\pi\Delta f t}. \quad (6)$$

This frequency error, $\Delta f$, thus introduces a linear phase ramping factor, $e^{j2\pi\Delta f t}$, on the received baseband signal $x(t)$.

A matched-filter-based detector performs the cross-correlation function between the incoming baseband signal in (6) and its local copy $x(t)$

$$\lambda(\tau, \Delta f) = \mathcal{E}^{-1} \cdot \int_0^{T_{sym}} x(t) \cdot e^{j2\pi\Delta f t} \cdot x^*(t - \tau) \, dt, \quad (7)$$

where $\mathcal{E} = \int_0^{T_{sym}} |x(t)|^2 \, dt$, $\tau$ is the lag of the cross-correlation function, and $T_{sym}$ is the preamble symbol duration.

In the absence of a frequency error, i.e., $\Delta f = 0$, the maximum output of the correlator (or the correlation peak) occurs at $\tau = 0$, $\lambda(\tau = 0, \Delta f = 0) = 1$, resulting in the optimal detection performance (in the sense of maximum likelihood) as asserted by the well-established matched-filter detection theorem. While in the presence of a frequency error, $\Delta f \neq 0$, the phase ramping component, $e^{j2\pi\Delta f t}$, effectively creates a "mismatch" between the incoming signal and the local copy of the matched filter, breaking the very premise of the optimality of a matched-filter detector, thereby inevitably leading to a loss in detection energy, i.e., $\lambda(\tau = 0, \Delta f \neq 0) < 1$, and ergo a loss in detection performance. This phenomenon becomes more pronounced in complexity- and energy-constrained mMTC, where the frequency error is likely large—to the extent that the resultant mismatch fails a matched-filter detector. Degraded detection performance causes repeated transmissions and detection attempts, accelerating the depletion of the transceiver battery.

Rather than changing the detector to get around the frequency error issue, which often results in complex and yet non-optimal solutions [27]-[29], we cling to the simple (yet optimal) matched-filter framework and look for a waveform that is robust against the distortion induced by the frequency error such that the matched filter's simplicity and optimality are maximally preserved. To that end, let us take another look at (7).

From the Cauchy-Schwartz inequality [30], we have the following relationship,

$$|\lambda(\tau, \Delta f)| \leq \mathcal{E}^{-1} \cdot \sqrt{\int_0^{T_{sym}} \left| x(t) \cdot e^{j2\pi\Delta f t} \right|^2 dt} \cdot \sqrt{\int_0^{T_{sym}} \left| x^*(t - \tau) \right|^2 dt}. \quad (8)$$

It follows that

$$|\lambda(\tau, \Delta f)| \leq 1, \quad \forall \Delta f \text{ and } \forall x(t), \quad (9)$$

noting that

$$\sqrt{\int_0^{T_{sym}} \left| x(t) \cdot e^{j2\pi\Delta f t} \right|^2 dt} = \sqrt{\int_0^{T_{sym}} \left| x^*(t - \tau) \right|^2 dt} \triangleq \sqrt{\mathcal{E}}. \quad (10)$$

(8) signifies that the maximum correlation value is mathematically attainable in the presence of a frequency error, $\Delta f \neq 0$, which is further asserted by the Cauchy-Schwartz inequality stating that equality in (8) holds if and only if

$$x(t) \cdot e^{j2\pi\Delta f t} = c \cdot x(t - \tau), \quad (11)$$

where $\forall c \in \mathbb{C}$ is a nonzero constant but with a unity magnitude for (10) to hold, and $\mathbb{C}$ denotes the set of complex numbers. This assertion means that the matched-filter detector can reclaim the total incoming signal energy even if there is a mismatch (a frequency offset) between the incoming signal and the detector.

We maintain that the class of waveforms that share the following canonical form satisfies (11),

$$s_{a,b,\rho}(t) = \rho \cdot e^{j\pi(at^2 + 2bt)}, \quad a\,(\neq 0), b \in \mathbb{R}, \; \rho \in \mathbb{C}, \quad (12)$$

where $\mathbb{R}$ denotes the set of real numbers. A factor of two is present before $b$ simply because $b$ has a special physical meaning, as will become clear next.

When (12) is used as a preamble baseband signal, from (4), the resultant RF signal,

$$\tilde{s}(t) = s_{a,b,\rho}(t) \cdot e^{j2\pi f_c t}, \quad 0 \leq t < T_{sym}, \quad (13)$$

has a waveform with an instantaneous frequency shift of

$$\Delta f_{a,b}(t) \triangleq a \cdot t + b, \quad 0 \leq t < T_{sym}, \quad (14)$$

off the carrier frequency, $f_c$, where $b$ is a frequency that stays unchanged within the preamble symbol duration, $T_{sym}$ (sec), whereas the instantaneous frequency shifts continuously and linearly at a rate of $a$ per unity time (one sec); hence, the name *frequency spread factor* for $a$.

Since $\rho$ remains constant throughout the preamble symbol duration, the corresponding RF waveform $\tilde{s}(t)$ has a constant profile (or envelope), enjoying a 3-dB PAPR—the same as an *unmodulated* sinusoidal waveform, hence, the lowest PAPR for an RF signal—ideal for the RF power amplifier and, hence, highest energy efficiency [21].

It is understood that unlike in (6), wherein the frequency *error*, $\Delta f$, is due to the inevitable frequency mismatch present between practical transceivers; here, the frequency offset, $\Delta f_{a,b}(t)$, in (14) of the canonical waveform, particularly $a$ and $b$, is by design and known *a priori* to the receiver.

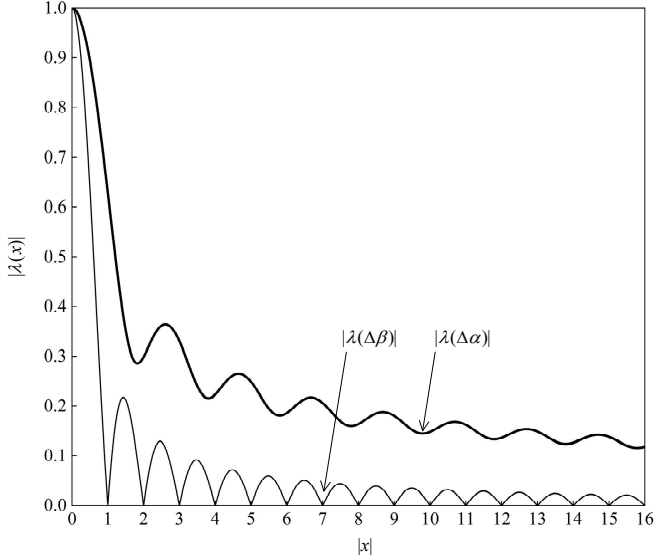

Figure 3 GM symbol waveform cross-correlations as a function of $\Delta\beta$ and $\Delta\alpha$ given in (24) and (27), respectively, to show the asymptotic nulls in the $\alpha$-dimension for potential $\alpha$-shift modulation as in the $\beta$-dimension.

The frequency error resilience and high energy amplifier efficiency of this class of waveforms render it an ideal candidate for mMTC, especially suitable for preamble detection with large initial frequency errors.

### B. Traffic Signal for MTC

A traffic signal is an information-bearing waveform of a physical layer transmission burst for carrying upper-layer data traffic (e.g., application traffic). It is assumed that the communicating ends at this stage are time- and frequency-synchronized (with some residual errors) owing to the preamble at the beginning of the transmission burst discussed above (Figure 2). Therefore, the purpose and operating conditions of the waveforms used for synchronization and data traffic are distinctive, leading to significantly different characteristics and design requirements for these two types of waveforms.

Nonetheless, in the following, we show that the dimensions from the canonical waveform used in the preamble waveform for resilience to frequency errors can also be exploited for conveying information and resilience to channel impairments. This is because, upon the preamble detection, the frequency offset has been corrected, and the dimensions are thus freed. Furthermore, this canonical waveform provides a *unified* waveform for preamble *and* data transmission, meaning the same waveform can be used for the whole transmission burst to reduce transceiver complexity and cost, which is particularly beneficial to mIoT devices.

For generality and notational simplification, we normalize time $t$ (sec) in (12) with respect to the symbol duration, $T_{sym}$ (sec), by letting

$$t \triangleq t/T_{sym} = t \cdot w_{sym}, \quad (15)$$

where $w_{sym} \triangleq T_{sym}^{-1}$ (Hz) is the symbol rate. It leads to a time-normalized (to symbol duration) canonical baseband waveform from (12),

$$s_{\alpha,\beta,\rho}(t) = \rho \cdot e^{j\pi(\alpha t^2 + 2\beta t)}, \quad 0 \le t < 1 \quad (16)$$
$$= \rho \cdot e^{j\pi(\alpha t^2 + 2\beta t)} \cdot g(t),$$

where $\alpha \triangleq a/w_{sym}^2$, $\beta \triangleq b/w_{sym}$, i.e., the scaled versions of $a$ and $b$ in (12), and $g(t)$ is the rectangular or gate function. Note that the constraint of $\alpha = 0$ in (12) no longer applies since the $\alpha$-dimension is now used for achieving different goals, and new constraints will be derived.

A closer look at (16) reveals that this canonical waveform encompasses three *independent* parameters, $\rho$, $\alpha$, and $\beta$, which are three dimensions that can potentially be exploited for carrying traffic under the most generalized framework.

Specifically, we wish that in a given time interval, 1 or $T_{sym}$ (sec), (16) carries $\ell_{sym}$ bits of information:

$$\mathbf{b} \triangleq [b^1 \ b^2 \cdots b^{\ell_{sym}}], \quad \ell_{sym} \ge 1. \quad (17)$$

To this end, we constitute the following generalized modulation or GM symbol set in the space spanned by $\rho$, $\alpha$, and $\beta$ provided by (16),

$$\mathcal{S}^{GM} \triangleq \left\{ s_{\alpha,\beta,\rho}(t) \middle| \alpha \in \mathcal{A}^{GM}, \beta \in \mathcal{B}^{GM}, \rho \in \mathcal{C}^{GM} \right\}, \quad (18)$$

where $\mathcal{A}^{GM}, \mathcal{B}^{GM} \subset \mathbb{R}$ and $\mathcal{C}^{GM} \subset \mathbb{C}$. The number of information bits a symbol carries, i.e., the *symbol payload*, is thus

$$\ell_{sym} = \left\lfloor \log\left(N \cdot M \cdot L\right) \right\rfloor \quad (19)$$
$$= \left\lfloor \log N + \log M + \log L \right\rfloor,$$

where

$$N \triangleq |\mathcal{A}^{GM}|, \ M \triangleq |\mathcal{B}^{GM}|, \ \text{and} \ L \triangleq |\mathcal{C}^{GM}|, \quad (20)$$

are the cardinalities of the waveform sets in the $\alpha$, $\beta$, and $\rho$ dimensions and, henceforth, coined *modulation orders* in the corresponding dimensions. Apparently,

$$\ell_{sym} \ge n + m + l, \quad (21)$$

where

$$n \triangleq \left\lfloor \log N \right\rfloor, \ m \triangleq \left\lfloor \log M \right\rfloor, \ \text{and} \ l \triangleq \left\lfloor \log L \right\rfloor. \quad (22)$$

In this paper, $\ell_{sym}$ is in the unit of *bit*, and, hence, a logarithm (log) is being assumed to have a base of 2.

#### 1) Signaling over $\rho$

Since $\rho$ is a parameter that controls the initial phase and amplitude of a GM symbol of (16), the signaling scheme over the $\rho$-dimension is readily available following the classic linear modulation QAM concept, including PSK, under which $\rho \in \mathcal{C}^{GM}$, and $\mathcal{C}^{GM}$ is a predefined QAM constellation, i.e.,

$$\mathcal{C}^{GM} = \left\{ \mu_\kappa \cdot e^{j/2\pi\vartheta_\kappa} \ \middle| \ (\mu_\kappa, \vartheta_\kappa) \in \mathbb{R}^2, \ \kappa = 0, 1, \cdots, L-1 \right\}. \quad (23)$$

Each $\rho \in \mathcal{C}^{GM}$ corresponds to a *channel symbol* in the form of (13), providing a modulation order of $L$.

#### 2) Signaling over $\beta$

Similarly, for the $\beta$-dimension to provide a modulation order of $M$, we need $M$ waveforms modulated with $M$ distinct values of $\beta$. For best detection performance, we wish these waveforms are uncorrelated [21].

For two GM symbols (or two modulation waveforms) in the form of (16) with an offset of $\Delta\beta$ in the $\beta$-dimension, their cross-correlation can be shown to be

$$|\lambda(\Delta\beta)| = \frac{|\sin(\pi \cdot \Delta\beta)|}{|\pi \cdot \Delta\beta|}. \qquad (24)$$

It is apparent that (24) possesses a set of nulls at

$$\Delta\beta = \kappa, \quad \kappa \in \mathbb{Z} \setminus 0, \qquad (25)$$

where $\mathbb{Z}$ is the set of integers, meaning the canonical waveforms with an offset of $\kappa$ in $\beta$ are uncorrelated and orthogonal.

Enlightened by this property, we let

$$\beta \in \mathcal{B}^{\mathrm{GM}} \triangleq \{0, 1, \cdots, M-1\}, \qquad (26)$$

and (16) becomes an $M$-ary orthogonal $\beta$-shift keying.

### 3) Signaling over $\alpha$

As the-last-but-not-the-least parameter of the canonical waveform, $\alpha$ is a waveform parameter that controls the rate of frequency shifting *within* a symbol (whereas $\beta$ is the frequency shift *between* symbols).

Similarly to the $\beta$-dimension, we need a set of waveforms with distinct values of $\alpha$. To determine such $\alpha$ values, we examine the cross-correlation property of (18) with respect to $\alpha$, i.e., the cross-correlation between two GM symbols with an offset of $\Delta\alpha$ in the $\alpha$-dimension. It can be shown

$$|\lambda(\Delta\alpha)| = \frac{\left|\mathrm{F}\left(\sqrt{2|\Delta\alpha|}\right)\right|}{\sqrt{2|\Delta\alpha|}}, \qquad (27)$$

where F is the complex Fresnel Function. (27) is plotted in Figure 3 along with (24).

Unlike the sin function in (24) for the $\beta$-dimension, $\mathrm{F}(x)$ has no null at $x \neq 0$. Therefore, orthogonal modulation is not available in the $\alpha$-dimension in the strict sense. However, there exist "quasi-nulls" at

$$x \approx \pm 2\sqrt{z}, \quad z \in \mathbb{Z}_+, \qquad (28)$$

for $\mathrm{F}(x)$ and yet $\lambda(x) \to 0$, as $x \to \infty$, meaning the canonical waveforms with an $\alpha$ offset of

$$\Delta\alpha = 2z, \quad z \in \mathbb{Z} \setminus 0 \qquad (29)$$

in the $\alpha$-dimension are asymptotically uncorrelated or *asymptotically orthogonal*, as shown in Figure 3. It leads to asymptotically orthogonal $\alpha$-shift keying on a symbol set of

$$\mathcal{A}^{\mathrm{GM}} \triangleq \{\alpha \in \mathbb{R} \,|\, \Delta\alpha = 2z, z \in \mathbb{Z} \setminus 0\}. \qquad (30)$$

Similarly to signaling in the $\beta$-dimention, we need $N$ distinct $\alpha$'s for a modulation order of $N$. How to select them from (30) is deferred to the next section when more constraints are enforced on (30) to meet specific mMTC requirements.

In fact, even the nulls in the $\beta$-domain ($\mathcal{B}^{\mathrm{GM}}$) are no longer strict but asymptotic once the bandwidth constraint is applied, as will be seen next.

### 4) The Bandwidth Constraint

Given $\alpha \in \mathcal{A}^{\mathrm{GM}}$, $\beta \in \mathcal{B}^{\mathrm{GM}}$, and $\rho \in \mathcal{C}^{\mathrm{GM}}$, the corresponding instantaneous frequency offset of the GM symbol waveform, $s_{\alpha,\beta,\rho}(t)$, in (16) is

$$\Delta\upsilon_{\alpha,\beta}(t) \triangleq \alpha \cdot t + \beta, \quad 0 \leq t < 1, \qquad (31)$$

according to (14). It is understood that, unlike the frequency offset of the preamble waveform in (14) (a function of $a$ and $b$) that is known to the receiver, here, the frequency offset, $\Delta\upsilon_{\alpha,\beta}(t)$, is introduced via $\alpha$ and $\beta$ at the transmitter to convey information to the receiver and, therefore, unknown to the receiver (before a successful detection).

Under $\mathcal{B}^{\mathrm{GM}}$, $\mathcal{S}^{\mathrm{GM}}$ in (18) constitutes a set of $M$ orthogonal symbol waveforms in the $\beta$-dimension, uniformly spaced 1 or $w_{\mathrm{sym}}$ (Hz) apart at $|\mathcal{B}^{\mathrm{GM}}| = M$ discrete frequencies contained in $\mathcal{B}^{\mathrm{GM}}$, which requires a channel bandwidth of at least

$$\mathcal{W} \triangleq M \text{ (DoF/symbol)}, \qquad (32)$$

i.e., occupied resources per symbol, corresponding to a physical bandwidth of

$$W \triangleq w_{\mathrm{sym}} \cdot \mathcal{W} \text{ (Hz)}, \qquad (33)$$

according to (15). On top of that, according to (31), the frequency offset shifts continuously and linearly at a rate of $\alpha$ per unit time or $w_{\mathrm{sym}} \cdot \alpha$ (Hz) per symbol.

In practice, a frequency channel for communication is always bandwidth-limited. We thus confine the dynamic range of (14) within $W$ (Hz), corresponding to confining (31) within $\mathcal{W}$ through (33) by letting

$$\begin{aligned} \Delta\upsilon_{\alpha,\beta}(t) &= (\alpha t + \beta) \bmod \mathcal{W} - \mathcal{W}/2 \\ &= (\alpha t + \beta) \bmod M - M/2, \end{aligned} \qquad (34)$$

where "mod" is the modulo operator, so that

$$\Delta\upsilon_{\alpha,\beta}(t) \in [-\mathcal{W}/2, \mathcal{W}/2) = [-M/2, M/2), \quad \forall\beta \in \mathcal{B}^{\mathrm{GM}}. (35)$$

The choice of $W$ (Hz) is dependent on the available deployment channel bandwidth, $W_{\mathrm{pass}}$ (Hz). Therefore, given $W_{\mathrm{pass}}$ (and, hence, $W$), the larger the $\mathcal{W}$ (or $M$) is, the lower the $w_{\mathrm{sym}}$.

Under (34), we have

$$\int_0^t \Delta\upsilon_{\alpha,\beta}(\nu)\, d\nu = (\alpha/2)t^2 + (\beta - M/2)t + \gamma_{\alpha,\beta}(t)/2, (36)$$

where

$$\gamma_{\alpha,\beta}(t) \triangleq -2M \cdot \sum_{q \in \mathrm{Q}_{\alpha,\beta}} \left(t - t_{\alpha,\beta}^q\right) \cdot u(t - t_{\alpha,\beta}^q) \cdot \mathrm{sign}\,\alpha, \quad (37)$$

$\mathrm{sign}(\bullet)$ represents the sign function, and $u(\bullet)$ is the Heaviside step function which is a result of the modulo operation in (34),

$$t_{\alpha,\beta}^q = \begin{cases} (qM - \beta)/\alpha, & \alpha > 0, \\ ((1-q)M - \beta)/\alpha, & \alpha < 0, \\ \infty & \alpha = 0, \end{cases} \qquad (38)$$

and $\mathrm{Q}_{\alpha,\beta} \triangleq \left\{q \,\middle|\, 0 \leq t_{\alpha,\beta}^q < 1\right\}$. We assert that

$$Q_{\alpha,\beta} = \left| \mathbf{Q}_{\alpha,\beta} \right| = \left\| (\alpha+\beta)/M \right\|. \tag{39}$$

Under the bandwidth constraint imposed by (34), (16) is updated as

$$\begin{aligned} s_{\alpha,\beta,\rho}(t) &= \rho \cdot e^{j2\pi \int_0^t \Delta\nu_{\alpha,\beta}(\nu)d\nu} \cdot g(t) \\ &= \rho \cdot e^{j\pi\left(\alpha t^2 + (2\beta - M)t + \gamma_{\alpha,\beta}(t)\right)} \cdot g(t). \end{aligned} \tag{40}$$

We note that the bandwidth constraint invalidates (24), i.e., GM composed of symbol waveforms in (40) is no longer strictly orthogonal on the $\beta$-domain, $\mathcal{B}^{\text{GM}}$, given in (26). We defer this issue to the next section when we have a clear definition of GM for mMTC.

### 5) FSK, LoRa Modulation, and Others under GM

It is "generalized" not only because GM embraces both the preamble and traffic signaling but also because it encompasses the most common modulation schemes used in mMTC.

To see that, we consider a special case when $\rho$ assumes a QAM constellation, and $\alpha = \beta = 0$, under which (40) becomes the QAM symbol,

$$s_{\alpha=0,\beta=0,\rho}(t) = \rho, \quad \rho \in \mathcal{C}^{\text{GM}}. \tag{41}$$

When $\mu$ is further set to a constant, it becomes the PSK modulation, $\mathcal{C}^{\text{GM}}_{\mu=1}$, as used in NB-IoT (QPSK, i.e., $L = \left| \mathcal{C}^{\text{GM}}_{\mu=1} \right| = 4$).

In another case when $\rho$ is fixed and $\alpha = 0$, (40) is the legacy FSK,

$$s_{\alpha=0,\beta,\rho=1}(t) = e^{j\pi(2\beta-M)t} \cdot g(t), \quad \beta \in \mathcal{B}^{\text{GM}}. \tag{42}$$

Now, let $\alpha = M$, (40) becomes the mainstream LoRa modulation mentioned in Section I,

$$s_{\alpha=M,\beta,\rho=1}(t) = e^{j\pi\left(Mt^2+(2\beta-M)t+\gamma_{M,\beta}(t)\right)} \cdot g(t), \quad \beta \in \mathcal{B}^{\text{GM}}, \tag{43}$$

where $\gamma_{M,\beta}(t) = -2M \cdot \left(t - t^1_{M,\beta}\right) \cdot u(t - t^1_{M,\beta})$, and $M = 128, \cdots, 4096$, as specified by the LoRa modulation.

Other variants include $\rho$ assuming a QAM constellation, and $\alpha = 0$, under which (40) becomes FQAM,

$$s_{\alpha=0,\beta,\rho}(t) = \rho \cdot e^{j\pi(2\beta-M)t} \cdot g(t), \quad \beta \in \mathcal{B}^{\text{GM}}, \quad \rho \in \mathcal{C}^{\text{GM}}, \tag{44}$$

whereas when $\rho$ assumes PSK, and $\alpha$ is fixed to $M$, (40) is simply PSK-LoRa,

$$s_{\alpha=M,\beta,\rho}(t) = \rho \cdot e^{j\pi\left(Mt^2+(2\beta-M)t+\gamma_{M,\beta}(t)\right)} \cdot g(t), \quad \beta \in \mathcal{B}^{\text{GM}}, \rho \in \mathcal{C}^{\text{GM}}_{\mu=1}. \tag{45}$$

It is now evident that, under the GM framework, QAM (including QPSK) only takes the $\rho$-dimension for bearing information, whereas FSK and LoRa modulation only utilize the $\beta$-dimension for bearing information with $\alpha$ fixed to 0 and $M$ (not information-bearing), respectively. FQAM and PSK-LoRa take both $\rho$- and $\beta$-dimensions. Therefore, under the GM framework, these modulation techniques are simply special configurations of GM, and yet they all have one thing in common: they leave the $\alpha$-dimension essentially unexploited.

### III. GM UNDER THE MMTC CONSTRAINTS

In the previous section, we derive a generalized modulation, or GM given in (40) from the canonical waveform, which possesses three independent parameters, constituting a signaling space spanned by $(\alpha, \beta, \rho)$. In this section, we *rethink* the mMTC modulation waveform design under the GM framework and examine how these parameters *individually* and *jointly* affect the overall modulation characteristics with regard to the mMTC requirements. The GM symbol waveform defined in (40) is thus further constrained to meet the requirements, which is, henceforth, referred to as mGM, i.e., "GM for mMTC."

### A. Energy Efficiency and Complexity Considerations

The first thing when it comes to mMTC design is the energy and implementation efficiency due to typically no mains supply and low-cost end devices. As touched upon before, a constant envelope waveform proves beneficial for both energy efficiency and the cost of the RF power amplifier of the transmitter [31]. Furthermore, memoryless modulation helps reduce transceiver complexity [32]. Continuous-phase favors spectral confinement [33] (details in B.2), which cancels out the need for spectral shaping that potentially raises the PAPR.

### 1) Energy and complexity constraints

The constant amplitude is guaranteed by enforcing $|\rho|$ in (23) to be constant, e.g.,

$$\rho \in \mathcal{C}^{\text{GM}}_{\mu=1}, \tag{46}$$

which precludes amplitude-shift keying (ASK), meaning PSK only.

*Memoryless* modulation requires the waveform of the current modulation symbol to be independent of the information carried by the previous symbols. This independency simplifies transceiver implementation, especially demodulation at the receiver.

Under the memoryless condition, the instantaneous phase of the $i$th modulation symbol waveform can be represented as

$$\phi(i+t) = \vartheta_i + \int_0^t \Delta\nu_{\alpha_i,\beta_i}(\nu)\,d\nu, \quad 0 \le t < 1, \tag{47}$$

where $\vartheta_i$ is defined in (23), and $i \in \mathbb{Z}$ is the symbol index. It is self-evident that the integration of $\Delta\nu_{\alpha_i,\beta_i}(t)$ over $t$ guarantees that $\phi(t)$ is continuous at any instant *within* a symbol, i.e., $0 \le t < 1$. Therefore, the condition for *continuous-phase* modulation boils down to the phase continuity *between* two consecutive symbols:

$$\phi(i) = \phi(i^-), \quad \forall i \in \mathbb{Z}_+, \tag{48}$$

where the left side is the starting phase of symbol $i$, and the right side is the phase at the end of the previous symbol (symbol $i-1$). From (47), we obtain

$$\phi(i) = \vartheta_i, \tag{49}$$

and

$$\phi(i^-) = \vartheta_{i-1} + \int_0^{1^-} \Delta\nu_{\alpha_{i-1},\beta_{i-1}}(\nu)\,d\nu. \tag{50}$$

Therefore, (48) implies

$$\vartheta_i = \vartheta_{i-1} + \int_0^{\Gamma^-} \Delta \upsilon_{\alpha_{i-1}, \beta_{i-1}}(\nu) \, d\nu, \quad \forall i \in \mathbb{Z}_+. \tag{51}$$

However, according to the memoryless constraint, the waveform of symbol $i$ cannot be affected by the information carried by symbol $i-1$. Consequently, (51) can only be satisfied by the following two conditions without breaching the memoryless constraint:

First, $\vartheta$ cannot be information-dependent; hence, $\vartheta \in \mathbb{R}$ is a constant, which precludes PSK. (46) then becomes

$$\mathcal{C}^{mGM} \triangleq \mathcal{C}_{\mu=1, \vartheta=0}^{GM}, \tag{52}$$

i.e., the $\rho$-dimension of GM is *precluded* from bearing information under this continuous-phase constraint, implying FQAM and PSK-LoRa are excluded from mGM.

Second, we require

$$\int_0^{\Gamma^-} \Delta \upsilon_{\alpha, \beta}(\nu) \, d\nu = z, \ z \in \mathbb{Z}, \ \forall \alpha \in \mathcal{A}^{GM}, \ \beta \in \mathcal{B}^{GM}, \tag{53}$$

where $\mathcal{B}^{GM}$ is defined in (26), and $\mathcal{A}^{GM}$, defined in (30), needs to be re-evaluated under (53) as follows.

From (36), given $M$, $\quad \forall \alpha \in \mathcal{A}^{GM}$ and $\beta \in \mathcal{B}^{GM}$, (53) becomes

$$\alpha/2 + \beta - M/2 + \gamma_{\alpha, \beta}(\Gamma^-)/2 = z, \ z \in \mathbb{Z}, \tag{54}$$

which further restricts $\alpha$ from $\mathcal{A}^{GM}$ to the following set:

$$
\begin{aligned}
\mathcal{A}^{mGM} \triangleq & \{0\} \\
& \cup \left\{ \alpha = \pm M/i \, \middle| \, \alpha = 2z, \ z \in \mathbb{Z} \backslash 0, i \in \mathbb{Z} \right\} \\
& \cup \left\{ \pm(i+1)M \, \middle| \, i \in \mathbb{Z}_+ \right\}.
\end{aligned}
\tag{55}
$$

We consider the $|\alpha| \leq M$ case, i.e.,

$$\mathcal{A}^{mGM} = \{0\} \cup \left\{ \alpha = \pm M/i \, \middle| \, \alpha = 2z, \ z \in \mathbb{Z} \backslash 0, i \in \mathbb{Z} \right\}. \tag{56}$$

We will justify the exclusion of $|\alpha| > M$ from mGM when we gain further insight into $\alpha$ in the subsequent sections.

The GM symbol waveform defined in (40) is thus further constrained in the domains, $\mathcal{A}^{mGM}$ and $\mathcal{C}^{mGM}$, to meet the energy efficiency and complexity constraints of mMTC. The corresponding modulation symbol set is thus,

$$\mathcal{S}^{mGM} \triangleq \left\{ s_{\alpha, \beta, \rho}(t) \, \middle| \, \alpha \in \mathcal{A}^{mGM}, \beta \in \mathcal{B}^{GM}, \rho \in \mathcal{C}^{mGM} \right\}. \tag{57}$$

Therefore, like LoRa modulation, mGM is memoryless and phase-continuous. Hence, mGM preserves both energy and implementation efficiency needed for mMTC.

### 2) Orthogonality Revisited

As stated earlier, the orthogonality of GM with respect to $\beta$ is invalidated as a result of the bandwidth constraint in (34) (recalling energy efficiency is affected by the orthogonality of the modulation waveforms through demodulation performance). We are now ready to revisit this issue under $\mathcal{A}^{mGM}$. To see how the orthogonality of GM on $\mathcal{B}^{GM}$ are affected by (34) under $\mathcal{A}^{mGM}$, given $M$, we divide $\mathcal{A}^{mGM}$ into two subsets with $\mathcal{A}_0^{mGM}$ corresponding to an $M$-ary orthogonal modulation set and $\mathcal{A}_\Gamma^{mGM}$ a set of non-orthogonal $M$-ary modulations.

Clearly, $\left\{ \alpha \in \mathcal{A}^{mGM} \, \middle| \, 0 \leq \alpha + \beta < M, \ \forall \beta \in \mathcal{B}^{GM} \right\}$ belongs to $\mathcal{A}_0^{mGM}$ since the bandwidth constraint has no effect on $\Delta \upsilon_{\alpha, \beta}$. More generally, it can be shown that

$$\mathcal{A}_0^{mGM} = \left\{ \alpha \in \mathcal{A}^{mGM} \, \middle| \, Q_{\alpha, \beta} = 0, \ \forall \beta \in \mathcal{B}^{GM} \right\}, \tag{58}$$

and

$$\mathcal{A}_\Gamma^{mGM} \triangleq \mathcal{A}^{mGM} - \mathcal{A}_0^{mGM}, \tag{59}$$

where $Q_{\alpha, \beta}$ is defined in (39), which yields

$$\mathcal{A}_0^{mGM} = \{0\}, \tag{60}$$

where $\alpha = 0$ corresponds to the $M$-ary FSK modulation set and

$$\mathcal{A}_\Gamma^{mGM} = \left\{ \alpha = \pm M/i \, \middle| \, \alpha = 2z, \ z \in \mathbb{Z} \backslash 0, i \in \mathbb{Z} \right\}. \tag{61}$$

(61) contains $\alpha = M$, which corresponds to the special case of LoRa modulation. Thus, LoRa modulation is *not* an orthogonal modulation, and $\alpha$ is fixed to $M$, i.e., non-information-bearing.

Nevertheless, it can be shown that the cross-correlation of the mGM symbol waveforms in the $\beta$-domain satisfies

$$|\lambda(\Delta \beta)| \leq \frac{|\sin(\pi \cdot \Delta \beta)|}{|\pi \cdot \Delta \beta|} + o, \tag{62}$$

where

$$o \triangleq \frac{M/|\alpha|}{\sqrt{2M} \cdot \sqrt{M/|\alpha|}}, \quad \alpha \in \mathcal{A}_\Gamma^{mGM}. \tag{63}$$

Since $o \to 0$ as $\alpha \to \infty$, (62) approaches (24). Therefore, $\mathcal{S}^{mGM}$ is an *asymptotically* orthogonal modulation set on the $\beta$-domain, $\mathcal{B}^{GM}$, meaning mGM closes to orthogonal modulation when $\alpha$ becomes large.

Similarly, given $\beta \in \mathcal{B}^{GM}$, the cross-correlation of mGM symbol waveforms in the $\alpha$-dimension satisfies

$$|\lambda(\Delta \alpha)| < \frac{3\sqrt{2}}{\sqrt{|\Delta \alpha|}}, \quad \alpha \in \mathcal{A}^{mGM}. \tag{64}$$

It approaches zero as $|\Delta \alpha| \to \infty$. Therefore, like the $\beta$-dimension, the $\alpha$-dimension may also be exploited for bearing information.

### B. Resource Efficiency Considerations

For the information-modulated GM traffic signal transmitted over a channel bandwidth of $W_{pass}$ (Hz) at an information rate of

$$r_b = \ell_{sym} \cdot w_{sym} \text{ (bits/sec)}, \tag{65}$$

the resource efficiency defined in (2) can be represented as

$$
\begin{aligned}
u \text{ or } \eta^{-1} &= r_b / W_{pass} \\
&= \ell_{sym} / \mathcal{W}_{pass} \text{ (bits/DoF)},
\end{aligned}
\tag{66}
$$

where

$$\mathcal{W}_{pass} \triangleq W_{pass} / w_{sym} \text{ (DoF/symbol)}, \tag{67}$$

following the convention used in (33).

Maximizing resource efficiency is thus ultimately 1) maximizing symbol payload, $\ell_{sym}$ (bits/symbol), and 2)

minimizing occupied resources per symbol, $\mathcal{W}_{\text{pass}}$ (DoF/symbol), same as maximizing $w_{\text{sym}}$ (symbols/sec) under a given physical channel bandwidth of $W_{\text{pass}}$ (Hz).

### 1) Joint Shift Keying and Symbol Payload

The GM symbol payload is given in (19). With signaling in the $\rho$-dimension out of the picture under mGM, we set our sights on the $\alpha$-dimension —the dimension that has not yet been exploited for additional payload by the existing technologies.

In the previous section, we have shown the asymptotic orthogonality property of the $\alpha$-dimension, which facilitates asymptotically orthogonal $2^{\ell_{\text{sym}}}$-ary *joint $\alpha$- and $\beta$-shift keying* (or simply *joint shift keying*), where

$$\ell_{\text{sym}} = \lfloor \log N + \log M \rfloor \geq n + m \qquad (68)$$

from (19).

For practical applications, $M = 2^m$. Under this constraint, (68) reduces to

$$\ell_{\text{sym}} = \lfloor \log N + m \rfloor = n + m, \qquad (69)$$

trading symbol payload or resource efficiency for implementation efficiency at the receiver (details in Section IV.A), and (56) becomes

$$\mathcal{A}^{\text{mGM}} = \{0\} \cup \{\alpha = \pm M/2^i \,|\, i=0,1,\cdots,m-1\}. \qquad (70)$$

Given $M$, (70) contains

$$N = 2m + 1 \qquad (71)$$

distinct $\alpha$'s, among which, $\bar{N} \triangleq 2^n$ out of them,

$$\bar{\mathcal{A}}^{\text{mGM}} \triangleq \begin{bmatrix} \alpha_0 & \alpha_1 & \cdots & \alpha_{\bar{N}-1} \end{bmatrix} \subseteq \mathcal{A}^{\text{mGM}}, \qquad (72)$$

facilitates $\bar{N}$-ary $\alpha$-shift keying, where

$$n = \lfloor \log(2m+1) \rfloor, \qquad (73)$$

which adds $n$ more information bits per symbol than $\beta$-shift keying only (e.g., FSK and LoRa modulation), corresponding to a

$$\frac{\lfloor \log(2m+1) \rfloor}{m} \times 100\% \qquad (74)$$

gain in symbol payload ($\ell_{\text{sym}}$) from joint shift keying.

### 2) Spectral Characteristics and Symbol Rate

A second factor that affects the resource efficiency is the symbol rate, $w_{\text{sym}}$. The sustainable symbol rate for a given channel bandwidth of $W_{\text{pass}}$ is a quantity that is practically affected by the power spectral characteristics, particularly spectral confinement, of the symbol sequence. This point becomes clear once we have a clear picture of the power spectrum of a GM symbol sequence.

To that end, we derive the *power spectral density* of a sequence of the information-modulated GM symbols,

$$\sum_i s_{\alpha_i, \beta_i, \rho_i}(t - i), \qquad (75)$$

which can be shown to consist of two components,

$$P(\varsigma) = P^c(\varsigma) + P^d(\varsigma), \qquad (76)$$

where

$$P^c(\varsigma) = \text{Var}\left\{S_{\alpha,\beta,\rho}(\varsigma)\right\}, \qquad (77)$$

is the continuous component, and

$$P^d(\varsigma) = \left| \text{E}\left\{S_{\alpha,\beta,\rho}(\varsigma)\right\} \right|^2 \cdot \sum_{j=-\infty}^{\infty} \delta(\varsigma - j), \qquad (78)$$

is the discrete component (attributable to symbol periodicity) and $\delta(\bullet)$ is the Dirac delta function. $S_{\alpha,\beta,\rho}(\varsigma)$ is the Fourier Transform of the baseband modulation symbol $s_{\alpha,\beta,\rho}(t)$, and $\alpha$, $\beta$, and $\rho$ are independent random variables (functions of the information bits). Given $M$ and $\forall \beta \in \mathcal{B}^{\text{GM}}$, the probability of $\beta = \beta$ is $1/M$.

From the power spectral density, $P(\varsigma)$, the power spectrum of GM is obtained,

$$\bar{P}(\varsigma) = \sum_{j=-\infty}^{\infty} P_j \cdot g(\varsigma - j + 1/2), \qquad (79)$$

where

$$P_j = \int_{j-1/2}^{j+1/2} P(\varsigma) \, d\varsigma. \qquad (80)$$

In the previous section, the $\rho$-dimension is precluded from mGM for phase continuity due to the spectral confinement issue, and the justification is deterred. With the derived spectrum, we are ready to revisit this issue from the spectral characteristic perspective to see how phase discontinuity affects spectral confinement. We then examine the effect of the $\alpha$ dynamics on the spectral confinement to determine if the $\alpha$-dimension can be included in mGM.

When $\rho$ is involved in the modulation, $\rho \in \mathcal{C}^{\text{GM}}_{\mu=1}$ a PSK constellation defined in (46) with $\mu = 1$ and $\text{E}\{\rho\} = 0$, under which,

$$\text{E}\left\{S_{\alpha,\beta,\rho}(\varsigma)\right\} = \text{E}\{\rho\} \cdot \text{E}\left\{S_{\alpha,\beta}(\varsigma)\right\} = 0, \qquad (81)$$

and

$$\text{Var}\left\{S_{\alpha,\beta,\rho}(\varsigma)\right\} = \text{Var}\left\{S_{\alpha,\beta}(\varsigma)\right\} + \left|\text{E}\left\{S_{\alpha,\beta}(\varsigma)\right\}\right|^2, \qquad (82)$$

where $\alpha \in \mathcal{A}^{\text{mGM}}$ and $\beta \in \mathcal{B}^{\text{GM}}$, given $M$. According to (76), the power spectral density with $\rho \in \mathcal{C}^{\text{GM}}_{\mu=1}$ is

$$P_\vartheta(\varsigma) = \text{Var}\left\{S_{\alpha,\beta}(\varsigma)\right\} + \left|\text{E}\left\{S_{\alpha,\beta}(\varsigma)\right\}\right|^2, \qquad (83)$$

whereas the power spectral density with $\rho \in \mathcal{C}^{\text{mGM}}$ is

$$P^{\text{mGM}}(\varsigma) = \text{Var}\left\{S_{\alpha,\beta}(\varsigma)\right\} + \left|\text{E}\left\{S_{\alpha,\beta}(\varsigma)\right\}\right|^2 \sum_{j=-\infty}^{\infty} \delta(\varsigma - j). \qquad (84)$$

Figure 4 plots the power spectrum, $\bar{P}_\vartheta(\varsigma)$ and $\bar{P}^{\text{mGM}}(\varsigma)$, according to (79) under (83) and (84), where $\mathcal{C}^{\text{GM}}_{\mu=1}$ is a QPSK constellation, i.e., $L = \left|\mathcal{C}^{\text{GM}}_{\mu=1}\right| = 4$.

We observe that although the frequency deviation in (31) is restricted by $\mathcal{W}$ (given $M$), the actual spectral power spreads beyond $\mathcal{W}$. In the sequel, the frequency localization or spectral confinement is thus quantitatively measured by the percentage of the power outside $\mathcal{W}$, denoted as *OoW*. The OoW is associated with the dynamics *within* a symbol and *between* symbols. The former is affected by the value of $\alpha$ (i.e., how fast the frequency shifts *within* a symbol, given $M$) and, hence, termed the *intrinsic* OoW; the latter is attributable to changes of $\alpha$ *between* symbols (due to $\alpha$-shift keying) and phase discontinuity between symbols (if any) and coined the *extrinsic* OoW. Figure 4 clearly demonstrates the phase discontinuity introduced by QPSK ($L=4$) creates excessive extrinsic OoW.

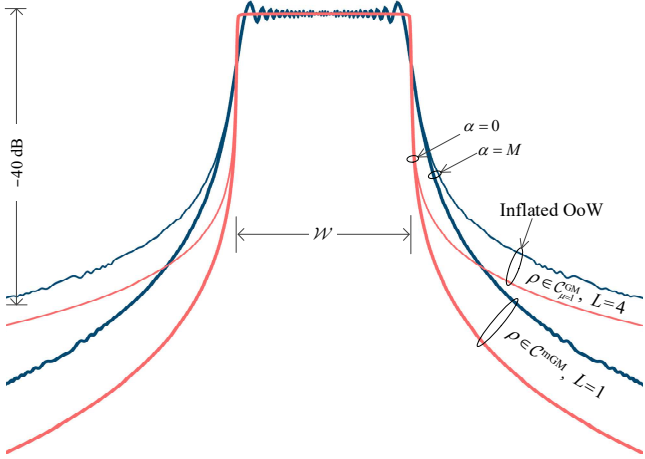

Figure 4 Power spectrum of GM, $\bar{P}^{\text{GM}}(\varsigma)$, for $L=1$ and $L=4$ (QPSK, thin lines) to show the effect of $\rho$ on spectral confinement, i.e., the out-of-$\mathcal{W}$ power (OoW), for $\alpha=0$, and $M$ (given $M=128$), under the same DoF.

For mGM, $\rho$-dimension is not used in the current design, the random variable $\boldsymbol{\rho}$ in (77) and (78) is fixed to a constant and can be omitted. Figure 5 plots the power spectrum of the baseline mGM configurations, i.e., with $M$-ary $\beta$-shift keying only, given $M=128$ ($m=7$). In addition, the power spectrum of $\bar{N}$-ary $\alpha$-shift keying on top of the baseline configurations (i.e., $\bar{N}$-ary $\alpha$-shift keying combined with $M$-ary $\beta$-shift keying) is plotted, where $\bar{N}=8$ ($n=3$) from (73).

We observe that, given $M$, a larger $|\alpha|$ corresponds to higher intrinsic OoW, (since the faster the frequency changes within a symbol becomes, the higher frequency it incurs and the larger the OoW). But what is worth noting is that $\bar{N}$-ary $\alpha$-shift keying causes only moderate growth in overall OoW power *relative to* $\alpha=0$—far less than $\alpha=M$ (LoRa) and $M/2$, by selecting $\bar{N}=8$ small-valued $\alpha$'s from $\left|\mathcal{A}^{\text{mGM}}\right|=15$ candidates [cf. (71)].

Now we have seen how different configurations of GM affect the spectral confinement or OoW of the modulation signals. To see how OoW has a direct effect on the sustainable symbol rate, $w_{\text{sym}}$ for a given bandwidth, $W_{\text{pass}}$, we refer to Figure 6, in which the bandwidth is practically specified by the power emission mask of the operating frequency band for a specific deployment [34].

In Figure 6, $W$ or the symbol rates ($w_{\text{sym}}$) for different configurations are scaled and maximized against the power spectral mask to show the impact of the spectral characteristics (i.e., OoW) on $W$ (or $w_{\text{sym}}$). Alternatively, instead of $W$, the transmit power can be scaled accordingly against the mask while maintaining the same $W$ across different configurations.

TABLE II summarizes the maximum symbol rate and information rate under various mGM configurations demonstrating the effect of signaling over the $\alpha$-dimension on the resource efficiency.

Under the same mask, TABLE III summarizes the sustainable symbol rate and information rate to show the effect of $\rho$-shift keying (QPSK). We see a significant negative effects on the resource efficiency because of excessive growth of the extrinsic OoW due to the phase discontinuity, which offsets the resource-efficiency benefit brought by signaling over the $\rho$-dimension. Spectrum shaping may improve the spectral confinement but escalates the carrier PAPR and compromises the energy efficiency. We thus see that using $\rho$-dimension to increase resource efficiency is ineffective in mMTC, hence, is not considered for mGM, i.e., $l = 0$. Instead, in the current paper, we focus on exploring the $\alpha$-dimension for carrying information, i.e., $\alpha$-shift keying, which is *not* used for bearing information in the existing modulation schemes, e.g., $\alpha$ is *fixed* (non-information-bearing) to 0 for FSK and $M$ for LoRa modulation.

TABLE II
SUSTAINABLE SYMBOL AND INFORMATION RATES OF mGM TO SHOW THE EFFECTS OF $\alpha$-SHIFT KEYING UNDER THE SPECTRAL MASK CONSTRAINT.

| $M$ value | $M=8$ | | $M=128$ | | $M=4096$ | |
|---|---|---|---|---|---|---|
| With $\alpha$-shift keying | No | Yes | No | Yes | No | Yes |
| Bits conveyed via $\alpha$-dimension, $n$ | 0 | 2 | 0 | 3 | 0 | 4 |
| Bits conveyed via $\beta$-dimension, $m$ | 3 | 3 | 7 | 7 | 12 | 12 |
| Symbol payload, $\ell_{sym}=m+n$ | 3 | 5 | 7 | 10 | 12 | 16 |
| Maximum symbol rate, $w_{sym}$ (symbols/sec) | 23,000 | 20,000 | 2,700 | 2,500 | 130 | 125 |
| Maximum information rate, $r_b$ (bits/sec) | 69,000 | 100,000 | 18,900 | 25,000 | 1,560 | 2,000 |

TABLE III
SUSTAINABLE SYMBOL AND INFORMATION RATES OF GM TO SHOW THE EFFECT OF $\rho$-SHIFT KEYING ($M=128$) UNDER THE SPECTRAL MASK CONSTRAINT.

| $|\alpha|$ value | $\alpha=0$ | | $|\alpha|=M/2$ | | $|\alpha|=M$ | |
|---|---|---|---|---|---|---|
| With $\rho$-shift keying | No | Yes | No | Yes | No | Yes |
| Bits conveyed via $\rho$-dimension, $l$ | 0 | 2 | 0 | 2 | 0 | 2 |
| Bits conveyed via $\beta$-dimension, $m$ | 7 | 7 | 7 | 7 | 7 | 7 |
| Symbol payload, $\ell_{sym}=m+l$ | 7 | 9 | 7 | 9 | 7 | 9 |
| Maximum symbol rate, $w_{sym}$ (symbols/sec) | 2,700 | 600 | 2,100 | 500 | 1,700 | 400 |
| Maximum information rate, $r_b$ (bits/sec) | 18,900 | 5,400 | 14,700 | 4,500 | 11,900 | 3,600 |

Compared with the $\rho$-dimension (cf. TABLE III), both boost symbol payload but, overall, $\alpha$-shift keying is spectrally more efficient. This justifies the inclusion of the new $\alpha$-domain ($\mathcal{A}^{\text{mGM}}$) signaling into mGM and preclusion of $\rho$-dimension from mGM. We thus rewrite (57) as

$$\mathcal{S}^{\text{mGM}} = \left\{ s_{\alpha,\beta}(t) \,\middle|\, \alpha \in \mathcal{A}^{\text{mGM}}, \, \beta \in \mathcal{B}^{\text{GM}} \right\}, \quad (85)$$

where

$$s_{\alpha,\beta}(t) = e^{j\pi\left(\alpha t^2 + (2\beta - M)t + \gamma_{\alpha,\beta}(t)\right)} \cdot g(t), \quad (86)$$

from (56).

Finally, one special case of $\alpha$-shift keying is worth mentioning: the *antipodal* binary $\alpha$-shift keying, under which (56) becomes

$$\alpha \in \bar{\mathcal{A}}^{\text{mGM}} = \left\{ \pm \alpha \,\middle|\, \forall \alpha \in \mathcal{A}_1^{\text{mGM}} \right\}, \quad (87)$$

which uses an antipodal pair of $\boldsymbol{\alpha} \in \{\pm \alpha\}$, given $\alpha \in \mathcal{A}_1^{\text{mGM}}$, to offer a modulation order of $N = 2$. (68) reduces to

$$\ell_{\text{sym}} = 1 + m. \quad (88)$$

It is self-evident that antipodal shift keying is not available for $\alpha \in \mathcal{A}_0^{\text{mGM}}$, i.e., 0.

Two unique properties of antipodal $\alpha$-shift keying merit attention.

First, we assert

$$\left| \mathrm{E}\left\{ S_{\alpha,\boldsymbol{\beta}}(\varsigma) \right\} \right|^2 = \left| \mathrm{E}\left\{ S_{-\alpha,\boldsymbol{\beta}}(\varsigma) \right\} \right|^2, \quad (89)$$

and

$$\mathrm{Var}\left\{ S_{\alpha,\boldsymbol{\beta}}(\varsigma) \right\} = \mathrm{Var}\left\{ S_{-\alpha,\boldsymbol{\beta}}(\varsigma) \right\}, \quad (90)$$

where $\boldsymbol{\beta} \in \mathcal{B}^{\text{GM}}$, given $\alpha \in \mathcal{A}_1^{\text{mGM}}$ and $M$. Consequently, both discrete and continuous components of the power spectral density are the same for $\pm\alpha$, i.e.,

$$P_\alpha^{\text{mGM}}(\varsigma) = P_{-\alpha}^{\text{mGM}}(\varsigma), \quad \forall \alpha \in \mathcal{A}_1^{\text{mGM}}, \, \beta \in \mathcal{B}^{\text{GM}}. \quad (91)$$

So is the corresponding power spectrum, $\bar{P}^{\text{mGM}}$, with and without antipodal $\alpha$-shift keying, meaning $\alpha$-shift keying for a given $\alpha$ does not change the power spectrum; hence, no changes to the sustainable symbol rate.

Second, antipodal binary $\alpha$-shift keying becomes relevant in the situation when only a particular $|\alpha|$ is adequate. As will be seen in the next section, mGM with $|\alpha| = M$ is most robust in a frequency-selective fading environment, and, therefore, antipodal binary $\alpha$-shift keying with $\boldsymbol{\alpha} \in \{\pm M\}$ provides the best resource efficiency and robustness.

### C. Robustness Considerations

As stated in the introduction, wireless communication is prone to low reliability due to channel impairments, mainly frequency-flat and frequency-selective fading, caused by multipath. Much research has been done in the past decades to address these impairments [35]-[38]. However, most of these solutions are for HTC, where cost and energy are of less concern. Enhancing communication robustness in situations with strict energy and cost constraints remains challenging. This section focuses on this topic, specifically, on how $\alpha$-dimension can be exploited for resilience to channel impairments without incurring a significant device cost penalty.

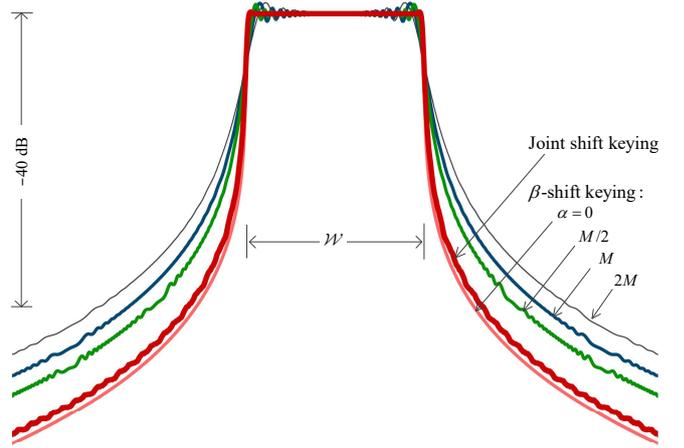

Figure 5 Power spectrum of mGM, $\bar{P}^{\text{mGM}}(\varsigma)$, plotted with different $\alpha$ configurations (given $M = 128$) to show the effect of $\alpha$ on OoW: 1) baseline configuration: $\beta$-shift keying only (i.e., $\alpha$ fixed) and 2) full-throttle configuration: joint $\alpha$- and $\beta$-shift keying, under the same DoF.

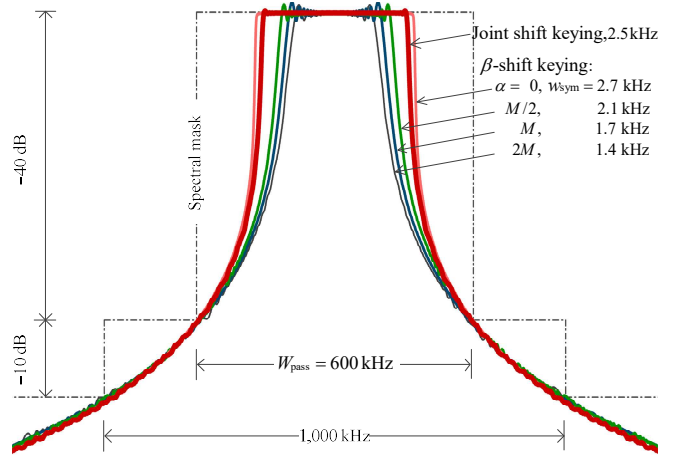

Figure 6 Power spectrum of mGM, $\bar{P}^{\text{mGM}}(f)$, where $f = \varsigma \cdot w_{\text{sym}}$ (Hz), plotted with different mGM configurations (given $M = 128$) against a spectral emission mask (deployment-dependent), with maximized $w_{\text{sym}}$ that satisfies the mask: 1) the baseline configuration: $\beta$-shift keying only (i.e., $\alpha$ fixed) and 2) the full-throttle configuration: joint $\alpha$- and $\beta$-shift keying.

In the above section, we find that a smaller $|\alpha|$ has better spectral confinement or less OoW. Therefore, a greater $|\alpha|$ consumes a larger bandwidth, and we must settle for a lower symbol rate. This raises the fundamental question of what is the point or benefit of having a larger or nonzero $\alpha$.

Since $\alpha$ is the linear frequency shift rate within a GM symbol, it is directly related to the bandwidth a symbol spreads and, hence, related to the *frequency diversity*. A larger $\alpha$ means more frequency diversity a symbol can benefit from available in a frequency-selective channel up to a point when $\alpha = M$, at which the frequency sweeps the entire bandwidth, $\mathcal{W}$ or $W$ (Hz), within a symbol according to (31), attaining the full frequency diversity. This aspect of $\alpha$ is relevant to mMTC in a wireless environment, which brings up another mMTC requirement, i.e., robustness. By robustness, we mean the ability to withhold channel impairments, particularly the ability to maintain a given required resource efficiency with a minimal cost in energy efficiency. We evaluate the choice of $\alpha$ on the robustness of mGM and explore the benefit of adapting $\alpha$ under different practical mMTC channels.

To look further into how $\alpha$ is related to the mGM's robustness in a practical wireless communication environment, we consider a general multipath channel represented by

$$\boldsymbol{h}(t) = \sum_i \boldsymbol{\rho}_i \cdot \delta(t - \boldsymbol{t}_i), \quad \boldsymbol{\rho}_i \in \mathbb{C}, \quad \boldsymbol{t}_i \in \mathbb{R}_+, \quad (92)$$

which captures the channel seen at the receiver, where $\mathrm{E}\{\boldsymbol{\rho}_i\} = 0$, $\mathrm{E}\{\boldsymbol{\rho}_i \cdot \boldsymbol{\rho}_j^*\} = \sigma_i \cdot \sigma_j \cdot \delta_{ij}$, and $\sum_i \sigma_i^2 = 1$ are assumed, and $\delta_{ij}$ is the Kronecker delta function.

Given $\boldsymbol{\rho}_i$, $\boldsymbol{t}_i$, and $\alpha \in \mathcal{A}^{\mathrm{mGM}}$, assume the mGM symbol of parameter $\beta \in \mathcal{B}^{\mathrm{GM}}$, i.e., $s_{\alpha,\beta}(t) \in \mathcal{S}^{\mathrm{mGM}}$, is transmitted (after up-conversion). Hereafter, we abuse the notation a bit and refer to it as the "$\beta$th symbol" for description simplicity. The signal arriving at the receiver (after down-conversion) is

$$\begin{aligned} \hat{s}_\beta(t) &\triangleq \left(s_{\alpha,\beta} \otimes h\right)(t) \\ &= \sum_i \rho_i \cdot s_{\alpha,\beta}(t - \boldsymbol{t}_i). \end{aligned} \quad (93)$$

The receiver consists of a bank of $M$ matched filters that matches to $\mathcal{S}^{\mathrm{mGM}}$ under channel (92), i.e.,

$$\left\{ g_\kappa(t) = \hat{s}_\kappa^*(1-t) \,\middle|\, \kappa \in \mathcal{B}^{\mathrm{GM}} \right\}, \quad (94)$$

of which, the output at the $\kappa$th filter is

$$\begin{aligned} y_{\kappa|\beta}(t) &= \left(\hat{s}_\beta \otimes g_\kappa\right)(t) \\ &= \sum_{i,j} \rho_i \cdot \rho_j^* \cdot \lambda_{\beta\kappa}(t, \boldsymbol{t}_j - \boldsymbol{t}_i), \end{aligned} \quad (95)$$

where

$$\begin{aligned} \lambda_{\beta\kappa}(t, \tau) &\triangleq \left\langle s_{\alpha,\beta}(t), s_{\alpha,\kappa}(t-\tau) \right\rangle \\ &= \int_0^1 s_{\alpha,\beta}(\nu) \cdot s_{\alpha,\kappa}^*(t - \tau - 1 + \nu) \, d\nu, \end{aligned} \quad (96)$$

is the cross-correlation between the $\beta$th and $\kappa$th symbols with a time offset $\tau$.

Accordingly, the output at the $\beta$th matched filter at time $t = 1^-$ (i.e., at the end of a symbol) is

$$y_{\beta|\beta}(1^-) = \sum_i |\rho_i|^2 + \underbrace{\sum_{i \neq j} \rho_i \cdot \rho_j^* \cdot \lambda_{\beta\beta}(1^-, \boldsymbol{t}_j - \boldsymbol{t}_i)}_{\text{leakage}}, \quad (97)$$

where $\lambda_{\beta\beta}(1^-, \tau)$ is the autocorrelation of $s_{\alpha,\beta}(t)$. Apparently, a nonzero $\lambda_{\beta\beta}(1^-, \tau)$ leads to energy "leakage" between paths at the matched filter output, causing fluctuations around

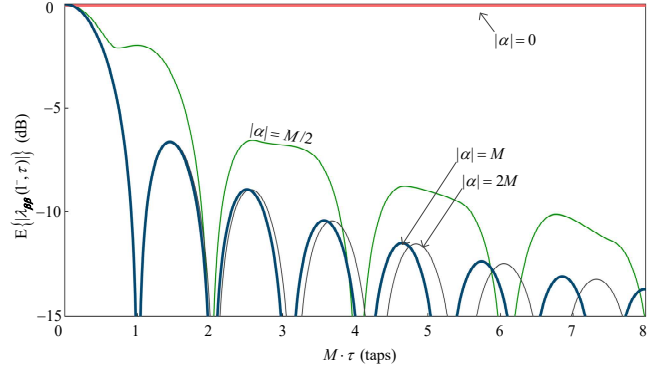

Figure 7 The autocorrelation of mGM symbols for various values of frequency spread factor, i.e., $\alpha$, under the same DoF given $M = 128$. One tap is defined as $\mathcal{W}^{-1}$.

$$\mathrm{E}\{\boldsymbol{y}_{\beta|\beta}(1^-)\} = \sum_i \mathrm{E}\{|\boldsymbol{\rho}_i|^2\} = 1, \quad (98)$$

resulting in "fading" in the detected energy of the $\beta$th matched filter, compromising the detection performance. The leakage that causes the fading can be measured as

$$\mathrm{Var}\left\{\sum_{i \neq j} \boldsymbol{\rho}_i \cdot \boldsymbol{\rho}_j^* \cdot \lambda_{\beta\beta}(1^-, \boldsymbol{t}_j - \boldsymbol{t}_i)\right\} = \sum_{i \neq j} |\lambda_{\beta\beta}(1^-, \boldsymbol{t}_j - \boldsymbol{t}_i)|^2 \cdot \sigma_i^2 \cdot \sigma_j^2, \quad (99)$$

which increases as $|\lambda_{\beta\beta}(1^-, \tau)|$ increases. It is thus clear that a larger autocorrelation causes larger fading at the output of the matched filter, decreasing the probability of successful detection.

Clearly, (92) is *frequency-flat* when $\mathrm{Var}\{\boldsymbol{t}_i\} \ll \mathcal{W}^{-2}$; it behaves increasingly *frequency-selective* when $\mathrm{Var}\{\boldsymbol{t}_i\}$ increases, and frequency-selective becomes predominant when $\mathrm{Var}\{\boldsymbol{t}_i\} > \mathcal{W}^{-2}$.

It is readily seen that

$$|\lambda_{\beta\beta}(1^-, \tau)| = 1, \quad \forall \tau, \ \alpha \in \mathcal{A}_0^{\mathrm{mGM}}, \ \beta \in \mathcal{B}^{\mathrm{GM}}, \quad (100)$$

whereas, the exact case for $\tau > 0$ and $\alpha \in \mathcal{A}_1^{\mathrm{mGM}}$ is mathematically cumbersome. Nevertheless, it can be shown that, given $M$,

$$|\lambda_{\beta\beta}(1^-, \tau)| < 1, \quad \forall \tau \neq 0, \ \alpha \in \mathcal{A}_1^{\mathrm{mGM}}, \ \beta \in \mathcal{B}^{\mathrm{GM}}. \quad (101)$$

Thereby, an mGM symbol with nonzero $\alpha$ suffers from less fading under a frequency-selective channel, meaning nonzero-$\alpha$ mGM is more resilient to frequency-selective fading.

From (100), an mGM symbol with $\alpha = 0$ has a maximum autocorrelation of 1 (0 dB) across *all* lags, therefore, is most vulnerable to frequency-selective fading. As plotted in Figure 7, as $|\alpha|$ increases (hence, the increased frequency spread), the symbol shows a growing trend of decreasing autocorrelation, reaching the minimum as $\alpha = M$, which benefits the symbol's resilience to frequency-selective fading the most. From (31), under $\alpha = M$, the frequency traverses the entire bandwidth ($\mathcal{W}$) over a symbol duration, enabling a receiver to capture the full frequency diversity the channel offers. This finding indicates that the $\alpha$-dimension provides *additional* leverage or degrees of freedom to control the robustness of mGM and adapt to various mMTC environments.

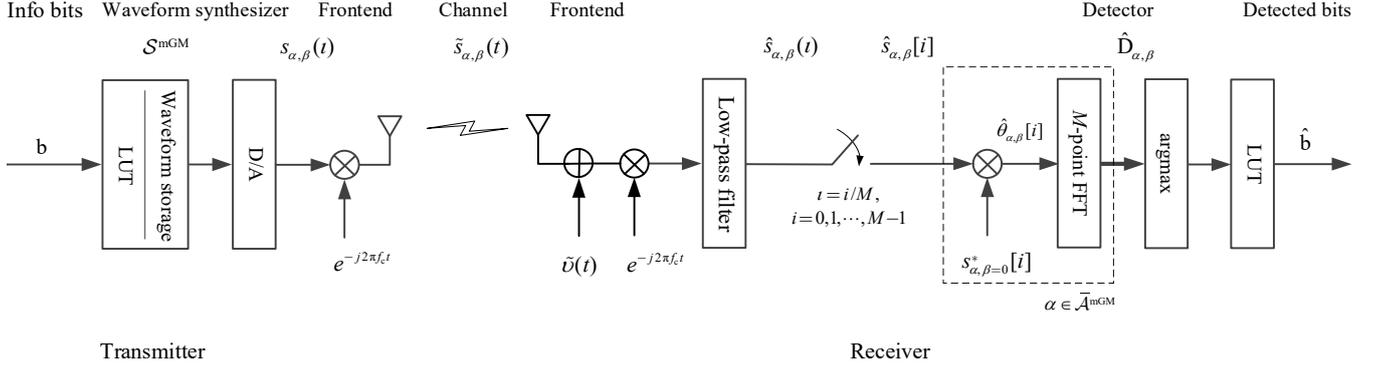

Figure 8 Abstracted practical LoRa transceiver structure (black) adapted for joint $\alpha$- and $\beta$-shift keying detection (red), where the $M$-point FFT-based detection operation (dashed) is repeated for each $\alpha \in \bar{\mathcal{A}}^{\mathrm{mGM}}$, i.e., $|\bar{\mathcal{A}}^{\mathrm{mGM}}|$ times, whereas $\bar{N}=1$ and $\alpha = M$ under LoRa. The frontend processors of the transceiver (e.g., the RF power amplifier at the transmitter, the low-noise RF amplifier and automatic gain control amplifiers, phase-locked loop, and voltage or. numerically controlled oscillator at the receiver) are not drawn. We also assume unit gain along the transmission and receiving chain to ease diagramming and discussion.

The trend discontinues as $|\alpha|$ becomes greater than $M$ (e.g., $|\alpha| = 2M$). Hence, an $\alpha$ larger than $M$ reaps no more robustness but only further aggravates the intrinsic OoW, justifying the exclusion of $|\alpha| > M$ from $\mathcal{A}^{\mathrm{mGM}}$ given in (56).

## IV. SIMULATION RESULTS

Equipped with the previous analytical results, in this section, we numerically evaluate the performance of the mGM defined in (85) under different channel conditions to validate the analytical results. The focus is on the effect of the uncovered $\alpha$-dimension on the mGM characteristics, mainly 1) energy and resource efficiency and 2) robustness to wireless channels, and seeking a practical modulation scheme for mMTC while maintaining the device simplicity.

### A. The Transceiver Structure

The transceiver structure is adopted from the LoRa transceiver commonly used in practice for its simplicity. Figure 8 is an abstract structural diagram, where the highlighted part (in red) in the receiver is due to the introduction of $\alpha$-shift keying.

The information bit sequence to be transmitted are grouped per the symbol payload, i.e., $\ell_{\mathrm{sym}} = m + n$ bits per mGM symbol from (69), where, $m$ is attributed to $M$-ary $\beta$-shift keying, and $n$ owing to $\bar{N}$-ary $\alpha$-shift keying. Each group ($\ell_{\mathrm{sym}}$ bits) is mapped to an mGM waveform, $s_{\alpha,\beta}(t) \in \mathcal{S}^{\mathrm{mGM}}$, defined in (85), using a look-up table (LUT). LUT contains a digitized version of $\mathcal{S}^{\mathrm{mGM}}$ stored in a ROS (read-only storage).

A dual DAC generates the in-phase (the real) and the quadrature-phase (the imaginary) components of the baseband analog components, $s_{\alpha,\beta}(t)$, $\alpha \in \mathcal{A}^{\mathrm{mGM}}$, $\beta \in \mathcal{B}^{\mathrm{GM}}$, and is up-converted into an RF signal, $\tilde{s}_{\alpha,\beta}(t)$, at the carrier frequency of $f_c$.

The receiver reverses the process by down-converting the incoming signal from RF back to baseband, which is low-pass filtered with a bandwidth of $W$ (or $\mathcal{W}$) and then sampled at the rate of $W$, where $W$ is defined in (33).

The sampled waveform after low-pass filtering is

$$\hat{\mathbf{s}}_{\alpha,\beta} \triangleq \left[ \hat{s}_{\alpha,\beta}[0] \ \hat{s}_{\alpha,\beta}[1] \cdots \hat{s}_{\alpha,\beta}[M-1] \right], \quad (102)$$

where $\alpha \in \mathcal{A}^{\mathrm{mGM}}$, $\beta \in \mathcal{B}^{\mathrm{GM}}$, and

$$\hat{s}_{\alpha,\beta}[i] \triangleq \hat{s}_{\alpha,\beta}(i\mathcal{W}^{-1}), \quad i = 0,1,\cdots,M-1. \quad (103)$$

Let us consider the baseband *signal path* in the receiver chain (i.e., introduced at the frontend) to see how FFT can be applied to the detection of $\hat{\mathbf{s}}_{\alpha,\beta}$. If we omit the low-pass filter for now so that OoW of the incoming signal, $s_{\alpha,\beta}(t)$, is preserved, i.e.,

$$\hat{s}_{\alpha,\beta}(t) = s_{\alpha,\beta}(t), \quad (104)$$

the element of (102) can be represented as

$$\hat{s}_{\alpha,\beta}[i] = s_{\alpha,\beta}[i], \ i = 0,1,\cdots,M-1, \quad (105)$$

where

$$\begin{aligned} s_{\alpha,\beta}[i] &= s_{\alpha,\beta}(i\mathcal{W}^{-1}) \\ &= e^{j\pi(\alpha(i/M)^2 + (2\beta \cdot M)i/M)} \\ &= s_{\alpha,0}[i] \cdot e^{j2\pi\beta i/M}. \end{aligned} \quad (106)$$

The input to the detector is then

$$\begin{aligned} \hat{\theta}_{\alpha,\beta}[i] &= \hat{s}_{\alpha,\beta}[i] \cdot s^*_{\alpha,0}[i] \\ &= e^{j\pi(\alpha-\alpha)(i/M)^2} \cdot e^{j2\pi\beta i/M}, \ i = 0,1,\cdots,M-1. \end{aligned} \quad (107)$$

Under LoRa, i.e., $\alpha = M$, (107) becomes

$$\hat{\theta}_{\alpha=M,\beta}[i] = e^{j2\pi\beta i/M}, \ i = 0,1,\cdots,M-1. \quad (108)$$

As such, (108) can be represented in the vector form as

$$\hat{\theta}_\beta = \Lambda_\alpha^{\mathrm{H}} \cdot \hat{s}_{\alpha,\beta}, \quad (109)$$

where $\Lambda_\alpha \triangleq \mathrm{diag}\{s_{\alpha,0}\}$ is an $M \times M$ diagonal matrix whose diagonal elements are made of $s_{\alpha,\beta=0}$, and

$$\hat{\theta}_\beta = \left[ 1 \ e^{j2\pi\beta/M} \cdots \ e^{j2\pi\beta(M-1)/M} \right]. \quad (110)$$

It is observed that $\hat{\theta}_\beta$ is Hermitian of the $\beta$th row of an $M \times M$ DFT matrix,

$$\mathbf{F} \triangleq \left[ \theta_0 \ \theta_1 \cdots \theta_\beta \cdots \theta_{M-1} \right]^{\mathrm{H}}, \quad (111)$$

i.e., $\hat{\theta}_{\beta} = \theta_{\beta}$, rendering the $M$-point DFT-based detector nothing but a matched-filter bank to $\left\{ \theta_{\beta} \middle| \beta \in \mathcal{B}^{\mathrm{GM}} \right\}$. Applying DFT to (109) leads to

$$\hat{d}_{\beta} \triangleq F \cdot \hat{\theta}_{\beta}^{\mathsf{T}} = d_{\beta}, \qquad (112)$$

where

$$d_{\beta} \triangleq F \cdot \theta_{\beta}^{\mathsf{T}} \qquad (113)$$

is a length-$M$, all-zero vector except the $\beta$th element.

In the presence of the low-pass filter, the removal of (intrinsic) OoW causes that (110) no longer holds. Therefore,

$$\hat{d}_{\beta} \neq d_{\beta}, \quad \forall \beta \in \mathcal{B}^{\mathrm{GM}}, \qquad (114)$$

because $\hat{\theta}_{\beta} \neq \theta_{\beta}$, and F no longer represents the matched-filter bank to $\left\{ \theta_{\beta} \middle| \beta \in \mathcal{B}^{\mathrm{GM}} \right\}$. This results in degraded detection performance and energy efficiency *in the presence of noise*.

Nonetheless, it can be shown that the intrinsic OoW diminishes as $M$ increases, so does the distortion of $\hat{s}_{\alpha,\beta}$ from the original $s_{\alpha,\beta}$ caused by the low-pass filter. Therefore,

$$\lim_{M \to \infty} \hat{\theta}_{\beta} = \theta_{\beta}, \qquad (115)$$

and, thus,

$$\lim_{M \to \infty} \hat{d}_{\beta} = d_{\beta}, \quad \forall \beta \in \mathcal{B}^{\mathrm{GM}}. \qquad (116)$$

The FFT-based detector in Figure 8 is thus an asymptotical maximum-likelihood detector, meaning the degradation diminishes as $M$ increases.

Under $(\bar{N} \times M)$-ary joint $\alpha$- and $\beta$-shift keying using the receiver structure in Figure 8, the $M$-point FFT-based detection operation is repeated $\bar{N}$ times, for each $\alpha \in \bar{\mathcal{A}}^{\mathrm{mGM}}$, denoted as $\alpha_0, \alpha_1, \cdots, \alpha_{\bar{N}-1}$, where $\bar{N} = \left| \bar{\mathcal{A}}^{\mathrm{mGM}} \right|$, and $\bar{\mathcal{A}}^{\mathrm{mGM}}$ is defined in (72).

As such, (109) extends to an $M \times \bar{N}$ matrix,

$$\hat{\Theta}_{\alpha,\beta} = \left[ \hat{\theta}_{\alpha_0,\beta} \ \hat{\theta}_{\alpha_1,\beta} \cdots \hat{\theta}_{\alpha,\beta} \cdots \hat{\theta}_{\alpha_{\bar{N}-1},\beta} \right]. \qquad (117)$$

Consequently, (112) becomes an $M \times \bar{N}$ matrix,

$$\hat{D}_{\alpha,\beta} \triangleq F \cdot \hat{\Theta}_{\alpha,\beta}^{\mathsf{T}}. \qquad (118)$$

And $\arg\max \left\{ \hat{D}_{\alpha,\beta} \right\}$ points to the $\beta$th-row and the $\alpha$th-column element.

For the most efficient FFT implementation, $M = 2^m$ or (69) is assumed. Compared to the case of $\beta$-shift keying only, i.e., $\bar{N} = 1$ (e.g., LoRa, under which $\alpha = M$), $\alpha$-shift keying incurs $\bar{N}-1$ additional detection operations—the price to pay for the $\bar{N}$ times higher modulation order. Nonetheless, as mentioned in the introduction, mIoT is characterized by short-burst data services with data flowing from devices (e.g., sensors) to the servers. Therefore, application traffic is uplink-dominant, under which the receiver is located at the control station (or gateway in LoRa terminology) of the mMTC infrastructure, where complexity and energy are less of a concern. Hence, $\alpha$-shift keying introduces an option for better satisfying the network capacity requirement for mIoT applications—an effective remedy for resource inefficiency for low-cost devices (like LoRa devices) that cannot afford advanced modulation and coding technologies.

### B. The Impact of the $\alpha$-Dimension

In the simulations, only the baseband part of the transceiver is simulated.

For the receiver implementation in Figure 8, the receiver is synchronized to the incoming traffic signal after a successful detection of the preamble of the transmission burst (cf. Figure 2), and the residual frequency and time errors are thus assumed negligible for the demodulation of the traffic signal.

Simulations are performed under an AWGN (or line-of-sight) channel and multipath channels. Under the multipath channels, $t_i$ is exponentially distributed with different values of $\mathrm{Var}\{t_i\}$. In one case, frequency-flat is dominant, and in another case, frequency-selective becomes predominant although the magnitude of selectivity may vary with the $W$ of a particular configuration (cf. Figure 6). We start with the mGM baseline configuration and then the full-throttle configuration to see what the effect of the $\alpha$-dimension has on modulation characteristics.

#### 1) mGM: The Baseline Configurations

Let us first consider the baseline mGM configurations, i.e., only the $\beta$-shift keying is used for conveying information, i.e., $n = 0$. Under the baseline configuration, $\alpha \in \mathcal{A}^{\mathrm{mGM}}$ is fixed for a given modulation order $M$. In the simulation, three configurations are considered: $\alpha = 0$ (i.e., FSK configuration), $\alpha = M/2$, and $\alpha = M$ (i.e., the LoRa modulation configuration).

As predicted from the previous analysis, mGM with a *smaller* $\alpha$ favors resource efficiency, and a larger $\alpha$ helps with resilience to frequency-selective fading. The results in Figure 9 confirm the characteristics, where the resource and energy efficiency are plotted in thin lines.

Under AWGN (or line-of-sight) channel, mGM with a smaller $|\alpha|$, shows better performance due to higher resource efficiency. And under a frequency-flat fading channel [16], the performance is similar but at a much higher energy cost due to fading. Specifically, the mGM configuration with $|\alpha| > 0$ is outperformed by the mGM with FSK configuration (i.e., $\alpha = 0$), mainly disadvantaged by the sustainable symbol rate due to poor spectral confinement (large OoW power at $|\alpha| > 0$).

While under frequency-selective channels, the results tell a quite different story. The impairment to mGM with a small $|\alpha|$ is detrimental rendering it unsuited for frequency-selective channels. In contrast, a larger $|\alpha|$ has better resistance to the impairments, indicating that the frequency diversity from the benefit of a large $|\alpha|$ helps combat frequency-selective fading more effectively.

Since mGM with different $\alpha \in \mathcal{A}^{\mathrm{mGM}}$ exhibits opposite effects on the resource efficiency and robustness characteristics, we run into a dilemma. On one side, for a given $M$, we want to make $|\alpha|$ as small as possible for the highest sustainable symbol rates or resource efficiency, while on the other side, it is just the opposite: we want to make $|\alpha|$ as large as possible to gain as much robustness as possible, i.e., least cost to energy efficiency for a given resource efficiency.

Now let us look at two special cases: From Section II.B, under the mGM framework, FSK corresponds to the mGM configuration with $\beta$-shift keying and $\alpha$ set to zero; LoRa modulation corresponds to $\beta$-shift keying with $\alpha$ configured to $M$. Therefore, the FSK configuration performs best, and the LoRa configuration the worst under line-of-sight and frequency-flat conditions. The positions are switched under frequency-selective channels. It is evident that any *fixed* $\alpha$ configuration, either 0 (FSK) or $M$ (LoRa modulation), is *not* the best for practical mMTC environments.

Motivated by the fact that mGM has a whole suite of $\alpha$, i.e., $\mathcal{A}^{\text{mGM}}$ defined in (56), to choose from, we make mGM *adapt* $\alpha$ to the deployment environment. For instance, a small $|\alpha|$ is more applicable for narrowband scenarios—most common in low-power mMTC (where channels are likely frequency-flat) [39] and for satellite-based mMTC (where channels are likely line-of-sight) [21]. A large $|\alpha|$ is more suitable for urban deployments, where frequency-selective channels are more likely [40], whereas an intermediate $|\alpha|$, e.g., $M/2$, is more effective in balancing robustness and efficiency in deployments with moderate frequency-selectivity, e.g., rural deployments [41].

### 2) mGM: The Full-Throttle Configurations

Now, let us consider the "full-throttle" mGM configuration, i.e., the case of mGM with joint $\alpha$- and $\beta$-shift keying activated. This configuration exploits additional degrees of freedom from the $\alpha$-dimension to facilitate $\alpha$-shift keying on top of the baseline configuration, engaging in both $\alpha$-shift and $\beta$-shift keying. Since the FSK and LoRa modulation configurations only facilitates $\beta$-shift keying (i.e., belonging to the baseline configuration), the full-throttle mGM configuration gives a boost over FSK and LoRa under all channel conditions.

Figure 9 (thick lines) shows the mGM performance with the full-throttle configuration. Under line-of-sight and frequency-flat conditions, through $\alpha$-shift keying, both resource and energy efficiency receive a significant boost. Since $|\alpha| < M$ is least favored under frequency-selective channels, the use of antipodal binary $\alpha$-shift keying (defined in (87) with $\alpha \in \{\pm M\}$) offers a clear resource- and energy-efficiency boost over the LoRa modulation configuration. We thus see that the degrees of freedom uncovered from the $\alpha$-dimension of mGM clearly demonstrate the effectiveness under all deployment conditions.

## V. CONCLUSION

Wireless communications are all about degrees of freedom that hold the key to almost all issues in wireless communications. However, it is also the most limited resource in wireless communications. Exploration for new dimensions under strict energy and cost constraints is thus the main focus of this paper. MTC for massive IoT (or mMTC) waveform design is a non-trivial task since the requirements work against each other: higher resource efficiency typically incurs higher complexity and lower energy efficiency unless new degrees of freedom are excavated. This paper does just that by deriving a generalized modulation for mMTC (or mGM), under which a new dimension is uncovered and exploited for the requirements.

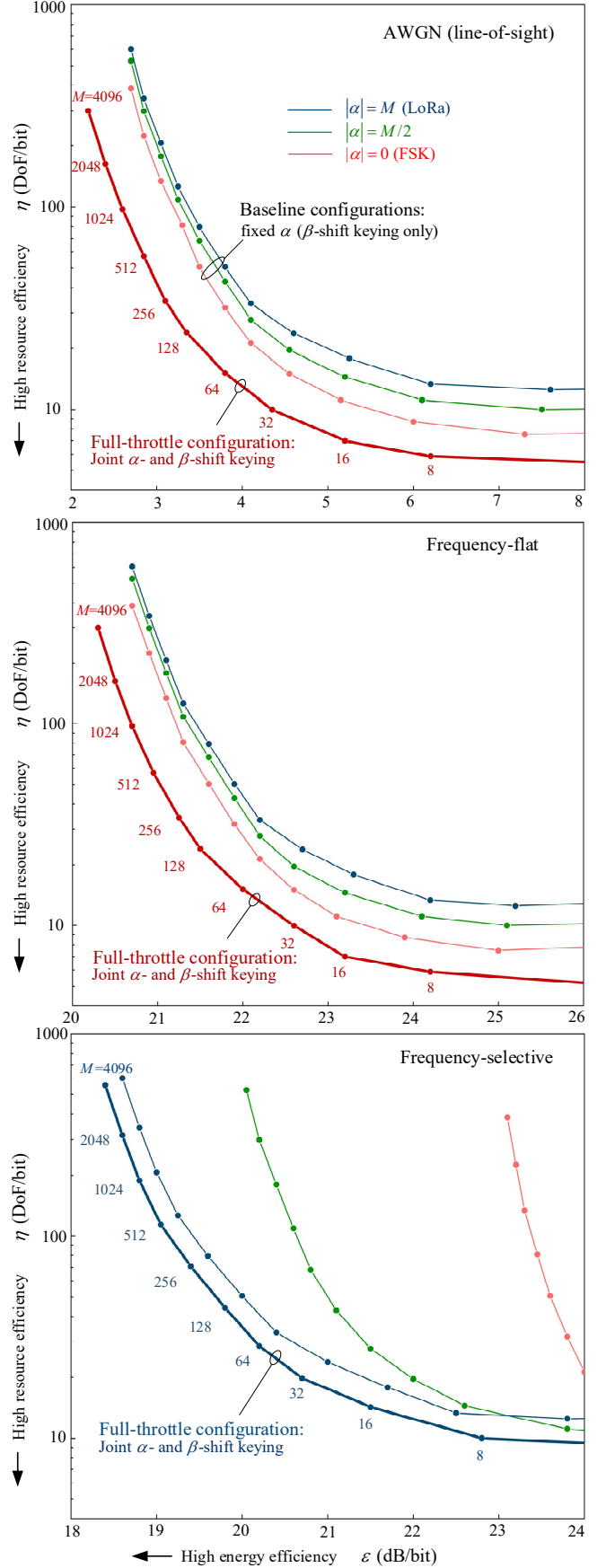

Figure 9 Energy and resource efficiency of mGM with baseline and full-throttle configurations and under different channel conditions.

Specifically, the dimensions from the canonical waveform are exploited *for different goals* during *different periods* of the transmission burst. During the preamble, the dimensions are utilized for 1) resilience to frequency error to address the low-frequency accuracy nature of a low-cost mIoT device oscillator. During the data transmission (during which the frequency ambiguity is resolved), the dimensions are re-utilized to improve 2) resource efficiency and 3) adaptability to channel conditions. As such, this canonical waveform provides 4) a *unified* waveform for the whole transmission burst, reducing overall transceiver complexity and cost.

The significance and unique contributions lie in that 1) it embraces various mMTC modulation schemes *under one unified framework*, enabling seamless adjustments or adaptation of modulation characteristics through configuration, and 2), more importantly, it opens up a new dimension that possesses some of the most desirable characteristics for energy-, resource-, and complexity-constrained mMTC applications and can be tailored for or adaptive to different deployment conditions. Specifically, we uncover and analyze the characteristics of the so-called $\alpha$-dimension from mGM. Instead of leaving it unexploited, we capitalize it as additional degrees of freedom to address the mMTC challenges: 1) resource and energy efficiency, 2) robustness against channel impairments, and 3) adaptivity to balance robustness and efficiency, all wrapped up in a single mGM framework. Finally, the mGM framework maximally retains the simplicity of LoRa devices for mIoT applications. The price is algorithmic complexity at the base station.

Immediate future work includes experimental validation of the proposed mGM focusing on the uplink under realistic channel conditions.